\documentclass[10pt,conference]{IEEEtran}
\IEEEoverridecommandlockouts

\usepackage{multirow}
\usepackage{subcaption}
\usepackage{enumitem}
\usepackage{algorithmic}
\usepackage{algorithm}
\usepackage{tcolorbox}
\usepackage{booktabs}
\usepackage{graphicx} 
\usepackage{bm}
\usepackage{xspace}
\usepackage{caption}
\usepackage{url}
\usepackage{amsmath,amssymb,amsfonts}
\usepackage{textcomp}
\usepackage{xcolor}
\usepackage{cite}
\usepackage{listings}  
\def\BibTeX{{\rm B\kern-.05em{\sc i\kern-.025em b}\kern-.08em
    T\kern-.1667em\lower.7ex\hbox{E}\kern-.125emX}}

\begin{document}

\title{What Makes Good In-context Demonstrations for Code Intelligence Tasks with LLMs?}

\author{\IEEEauthorblockN{Shuzheng Gao$^{1\dag}$, Xin-Cheng Wen$^{1}$, Cuiyun Gao$^{1\ast}$, Wenxuan Wang$^{2}$, Hongyu Zhang$^{3}$, Michael R. Lyu$^{2}$}

\IEEEauthorblockA{$^1$ School of Computer Science and Technology, Harbin Institute of Technology, Shenzhen, China}

\IEEEauthorblockA{$^2$ Department of Computer Science and Engineering, The Chinese University of Hong Kong, China}

\IEEEauthorblockA{$^3$ School of Big Data and Software Engineering, Chongqing University, China}

\IEEEauthorblockA{szgao98@gmail.com, xiamenwxc@foxmail.com, gaocuiyun@hit.edu.cn, hyzhang@cqu.edu.cn, \{wxwang,lyu\}@cse.cuhk.edu.hk}

\thanks{$^{\dag}$ The author is now affiliated with The Chinese University of Hong Kong.}
\thanks{$^{\ast}$ Corresponding author. The author is also affiliated with Peng Cheng Laboratory.}

}


\maketitle

\begin{abstract}
Pre-trained models of source code have gained widespread popularity in many code intelligence tasks. Recently, with the scaling of the model and corpus size, large language models have shown the ability of in-context learning (ICL).
ICL employs task instructions and a few examples as demonstrations, and then inputs the demonstrations to the language models for making predictions.
This new learning paradigm is training-free and has shown impressive performance in various natural language processing and code intelligence tasks.
However, the performance of ICL
heavily relies on the quality of demonstrations, e.g., the selected examples.
It is important to systematically investigate
how to construct a good demonstration for code-related tasks.
In this paper, 
we empirically explore the impact of three key factors on the performance of ICL
in code intelligence tasks: the selection, order, and number of demonstration examples. We conduct extensive experiments on three code intelligence tasks including code summarization, bug fixing, and program synthesis. Our experimental results demonstrate that all the above three factors dramatically impact the performance of ICL
in code intelligence tasks. Additionally, we summarize our findings and provide takeaway suggestions on how to construct effective demonstrations, taking into account these three perspectives. We also show that a carefully-designed demonstration based on our findings 
can lead to substantial improvements over widely-used demonstration construction methods, 
e.g., improving BLEU-4, EM, and EM by at least 9.90\%, 175.96\%, and 50.81\% on code summarization, bug fixing, and program synthesis, respectively.
\end{abstract}

\pagestyle{plain}

\maketitle

\section{Introduction}\label{sec:intro}
Recently, there has been an increasing focus on code intelligence research, aiming
at reducing the burden on software developers and enhancing programming productivity~\cite{DBLP:conf/issta/ZengTZLZZ22,DBLP:journals/corr/abs-2107-03374}. 
With the large-scale open-source code corpora and the progress of deep learning techniques, 
various 
neural source code models
have been developed and have achieved state-of-the-art performance on a variety of code intelligence tasks including code summarization~\cite{DBLP:conf/acl/AhmadCRC20}, bug fixing~\cite{DBLP:conf/icse/TufanoPWBP19}, and program synthesis~\cite{DBLP:conf/acl/YinN17}.

In recent years, the advent of pre-training techniques has significantly advanced progress in this area. For instance, 
CodeBERT~\cite{DBLP:conf/emnlp/FengGTDFGS0LJZ20}, a BERT-based model pre-trained on both natural and programming language data, has demonstrated promising performance in various code intelligence tasks~\cite{DBLP:conf/ijcai/HuLXLLJ18,DBLP:conf/icse/TufanoPWBP19}. Other subsequent pre-trained code models such as PLBART~\cite{DBLP:conf/naacl/AhmadCRC21} and CodeT5~\cite{DBLP:conf/emnlp/0034WJH21} further achieve much improvement over CodeBERT. However, the size and training data of the above models are limited, which may hinder the models from achieving their potential~\cite{DBLP:journals/corr/abs-2206-07682}.
In these years, we have witnessed explosive growth in the size of pre-trained models. Various billion-level large language models are proposed such as GPT-3~\cite{DBLP:conf/nips/BrownMRSKDNSSAA20} and PALM-E~\cite{DBLP:journals/corr/abs-2303-03378}. For instance, the size of the pre-trained
model PALM-E~\cite{DBLP:journals/corr/abs-2303-03378} (562B) in 2023 is over two thousand times larger than the largest model BERT~\cite{DBLP:conf/naacl/DevlinCLT19} (223M) in 2018.

\begin{figure}
    \centering
    \includegraphics[width=0.5\textwidth]{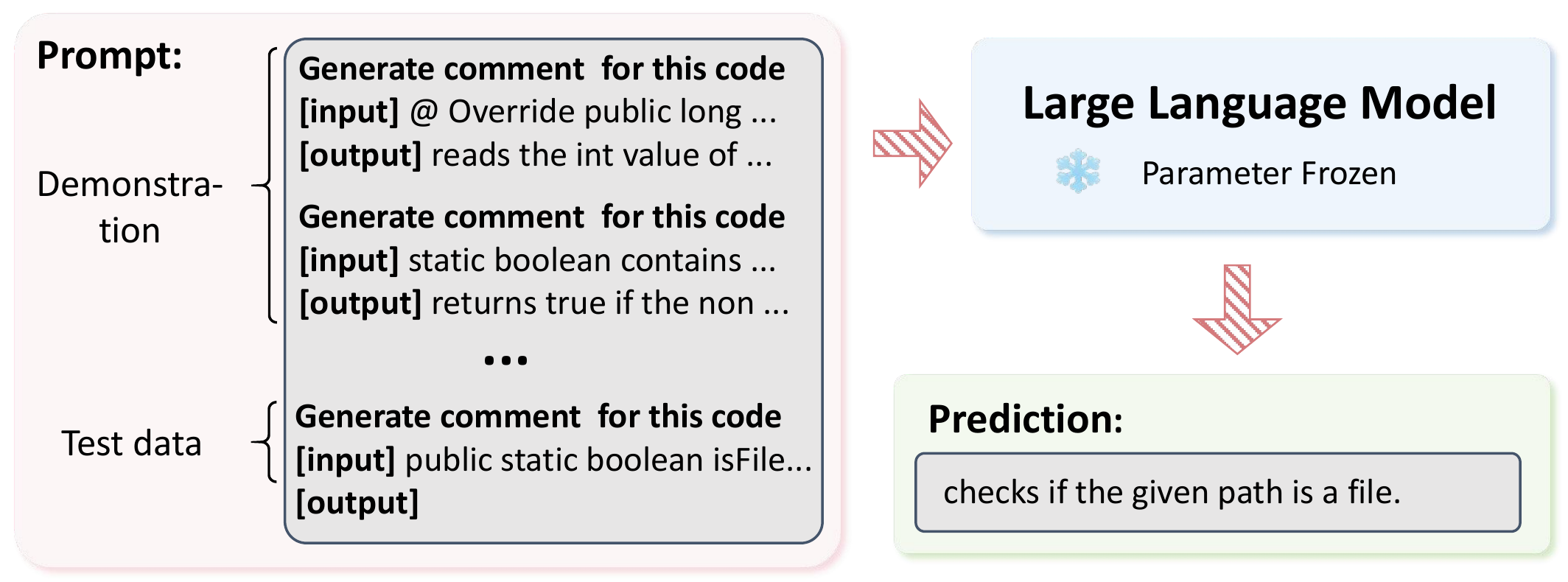}
    \caption{An example of in-context learning on code summarization task.}
    \label{fig:overview}
\end{figure}

As the size of language models and training data continues to increase, large language models (LLMs) demonstrate various emergent abilities. One such ability is in-context learning (ICL)~\cite{DBLP:conf/nips/BrownMRSKDNSSAA20,dong2022survey}, which allows models to learn from just a few examples within a specific context. As shown in Fig.~\ref{fig:overview}, ICL utilizes a demonstration including task instructions and a few examples to describe the task, 
which are then concatenated with a query question to form an input for the language model to make predictions. 
The most significant difference between ICL and traditional tuning methods such as fine-tuning~\cite{DBLP:conf/emnlp/FengGTDFGS0LJZ20} is that it is training-free and does not need 
parameter updates.
The training paradigm enables ICL to be directly used upon any LLMs and significantly reduces the training costs of adapting models to new tasks~\cite{DBLP:conf/nips/BrownMRSKDNSSAA20}.  
Recent studies show that ICL has achieved impressive results in various fields, including logic reasoning~\cite{DBLP:journals/corr/abs-2201-11903}, dialogue system~\cite{DBLP:conf/emnlp/Hu0X0SO22}, and program repair~\cite{DBLP:journals/corr/abs-2210-14179,DBLP:journals/corr/abs-2306-01394,DBLP:conf/icse-apr/PrennerBR22}, and can even outperform the supervised fine-tuning methods trained on
large task-specific data. 


Although ICL has been proven useful in code intelligence tasks, 
the performance of ICL is known to strongly rely on the quality of demonstrations~\cite{DBLP:conf/acl-deelio/LiuSZDCC22,DBLP:conf/acl/LuBM0S22}. Existing studies~\cite{DBLP:journals/corr/abs-2210-14179,DBLP:conf/icse-apr/PrennerBR22} mainly construct demonstrations by randomly selecting and arranging the demonstration examples.
To the best of our knowledge, there is a lack of an in-depth investigation of ICL for code intelligence tasks. 
Considering the impressive performance of ICL, it is worthy to understand the impact of demonstration design and investigate the challenges of applying ICL for code intelligence tasks.
In this work, we systematically analyze how different demonstration construction methods influence the performance of ICL on code intelligence tasks, aiming at answering the following question: \textbf{\textit{What makes good in-context demonstrations for code intelligence tasks with LLMs?}} By analyzing the design space of in-context demonstrations, our study mainly focuses on three aspects of in-context demonstrations, including the selection, order, and number of demonstration examples. 
We conduct an experimental study on three popular code intelligence tasks including code summarization, bug fixing, and program synthesis. Specifically, we mainly investigate the following four research questions (RQs):

\begin{enumerate}
    \item What kind of selection methods are helpful for ICL in code intelligence tasks?
    \item How should demonstration examples be arranged for ICL in code intelligence tasks?
    \item How does the number of demonstration examples in a prompt impact the performance of ICL in code intelligence tasks?
    \item How is the generalizability of our findings? 
\end{enumerate}

To answer the first RQ, we compare a wide range of demonstration selection methods such as random selection, similarity-based selection, and diversity-based selection. We also experiment with different retrieval methods in the similarity-based selection to
find which retrieval method is more
helpful for ICL in code intelligence tasks. To answer the second RQ, we compare random ordering with two other ordering methods including
similarity and reverse similarity, towards exploring the impact of different ordering methods.
To answer RQ3, we change the number of demonstration examples in the prompt and investigate whether the performance of ICL also grows with the increase in the number of demonstration examples. To answer the last RQ, we experiment on different LLMs and validate the findings we achieve in the above RQs.

\textbf{Key Findings.} Based on the extensive experiments, our study reveals several key findings: 

\begin{enumerate}
    \item Both similarity and diversity in demonstration selection are
    important factors for ICL in code intelligence tasks. They not only enhance the overall performance but also lead to more stable predictions.
    \item The order of demonstration examples has a large impact on the performance of ICL. 
    In most cases, placing similar samples at the end of a prompt achieves better results.
    \item Increasing the number of demonstration examples can be beneficial for ICL, provided that the examples are
    not cut off 
    due to the input length limitation of LLMs. Careful attention should be paid to this issue, as the length of code is generally longer than natural language.
\end{enumerate}

We also show that a  carefully-designed demonstration 
based on the achieved findings can lead to substantial improvements over the widely-used demonstration construction methods~\cite{DBLP:journals/corr/abs-2210-14179,DBLP:conf/icse-apr/PrennerBR22,DBLP:conf/kbse/Khan022}, e.g., improving BLEU-4, EM, and EM by at least 9.90\%, 175.96\%, and 50.81\% on code summarization, bug fixing and program synthesis, respectively.
 
\textbf{Contributions.} In summary, the main contributions of this work are as follows:
\begin{enumerate}
    \item To the best of our knowledge, this paper represents the first systematic study on how to construct effective demonstrations for code intelligence tasks.
    \item Our comprehensive exploration of demonstration design highlights a range of 
    findings for improving ICL's performance in code intelligence tasks.
    \item We discuss the implications of our findings for researchers and developers and future work for code intelligence tasks in the era of large language models.
\end{enumerate}


\section{Background}\label{sec:back}

\subsection{Large Language Models}\label{sec:back_llm}
LLMs have become a ubiquitous part of Natural Language Processing (NLP) due to their exceptional performance~\cite{DBLP:conf/nips/BrownMRSKDNSSAA20,GPT4}. These models typically follow the Transformer~\cite{DBLP:conf/nips/VaswaniSPUJGKP17} architecture and are trained on large-scale corpora using self-supervised objectives such as masked language modeling~\cite{DBLP:conf/naacl/DevlinCLT19}. The size of LLMs has increased significantly in the past few years. For example, the parameters of recent LLMs like GPT-3~\cite{DBLP:conf/nips/BrownMRSKDNSSAA20} and PALM-E~\cite{DBLP:journals/corr/abs-2303-03378} are over one hundred billion. 
Apart from the LLMs for general purposes, there are also LLMs with billion-level parameters trained on code corpora, such as 
AlphaCode~\cite{DBLP:journals/corr/abs-2203-07814}, and Codex~\cite{DBLP:journals/corr/abs-2107-03374}. 
The OpenAI's Codex is a large pre-trained code model that is capable of powering Copilot. AlphaCode~\cite{DBLP:journals/corr/abs-2203-07814} is a 41-billion-large model trained for generating code in programming competitions like Codeforces. Recently, LLMs like ChatGPT~\cite{ChatGPT} and GPT-4~\cite{GPT4} have also shown impressive performance in many code intelligence tasks.

Apart from proposing new LLMs, how to effectively leverage them has also become an important research topic. A prevalent method is to fine-tune the model and update its parameters on downstream datasets~\cite{DBLP:conf/naacl/DevlinCLT19}. Recently, prompt-based fine-tuning has been proposed, which aims to convert the training objective of downstream tasks into a similar form as the pre-training stage~\cite{DBLP:journals/csur/LiuYFJHN23,DBLP:conf/sigsoft/WangYGP0L22}. Considering the cost of tuning the whole model, various Parameter Efficient Tuning methods have been proposed, such as Adapter~\cite{DBLP:conf/icml/HoulsbyGJMLGAG19}, Lora~\cite{DBLP:conf/iclr/HuSWALWWC22}, and prefix tuning~\cite{DBLP:conf/acl/LiL20}. These methods keep most of the parameters in the model frozen and only tune a small portion of them. 

\begin{figure}
    \centering
    \includegraphics[width=0.47\textwidth]{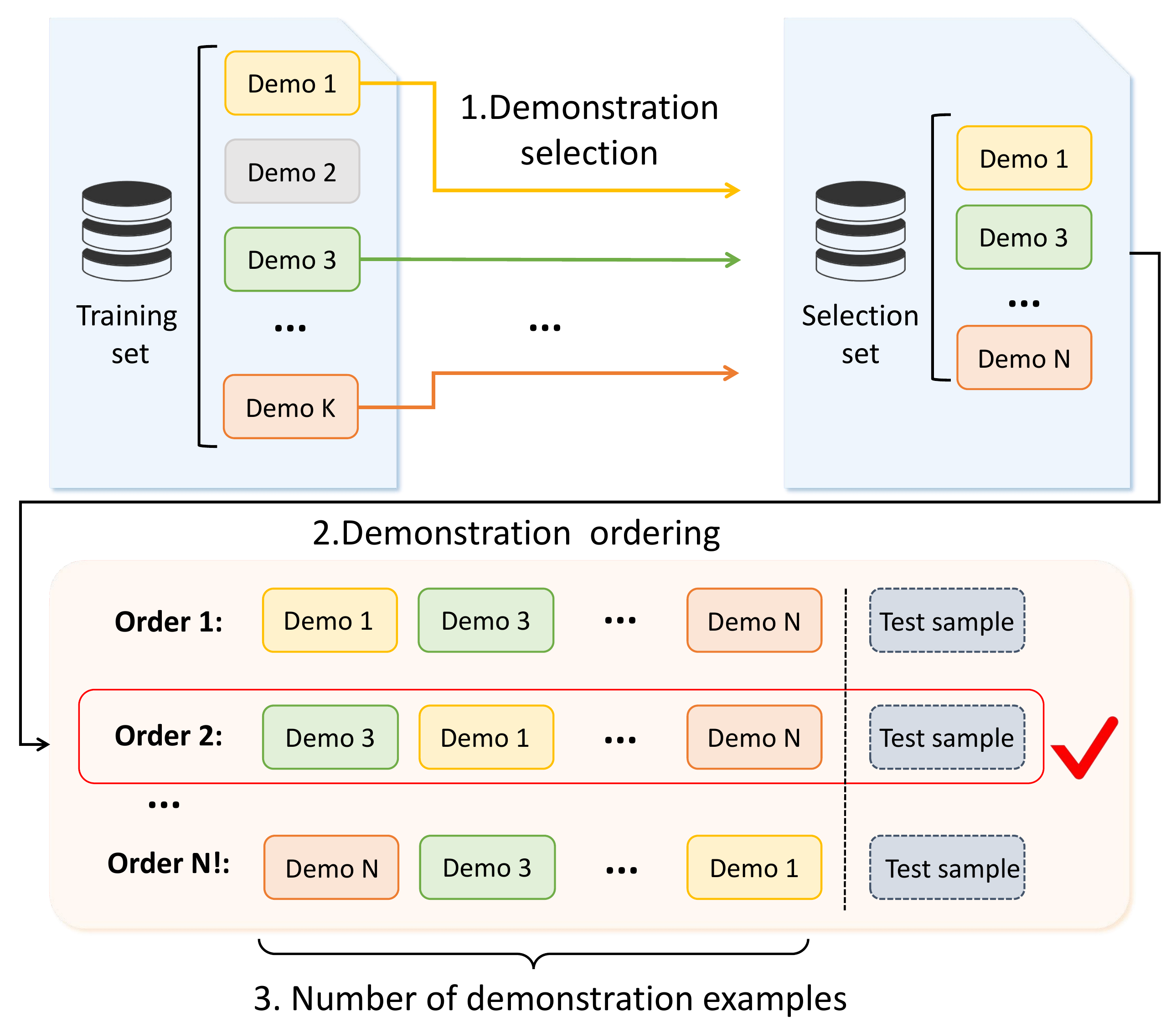}
    \caption{Illustration of design space of in-context demonstrations.}
    \label{fig:example}
\end{figure}


\subsection{In-context Learning}\label{sec:back_icl} 
Tuning a large pre-trained model can be expensive and impractical for researchers, especially when limited fine-tuned data is available for certain tasks. ICL offers a new alternative that uses language models to perform downstream tasks without requiring parameter updates~\cite{dong2022survey,DBLP:conf/nips/BrownMRSKDNSSAA20}. It leverages a demonstration in the prompt to help the model learn the input-output mapping of the task. 
This new paradigm has achieved impressive results in various tasks such as logic reasoning and program repair~\cite{DBLP:journals/corr/abs-2201-11903,DBLP:journals/corr/abs-2210-14179,DBLP:conf/icse-apr/PrennerBR22}.

Specifically, as shown in Fig.~\ref{fig:overview}, ICL employs $N$ demonstration examples $\{(x_1,y_1), (x_2,y_2),..., (x_N,y_N)\}$ and further reconstructs them into reconstructed examples $\{(x'_1,y'_1), (x'_2,y'_2),..., (x'_N,y'_N)\}$ by natural language instructions and prompt template, where $x_i$, $y_i$, $x'_i$, $y'_i$ are the input, output, reconstructed input, and reconstructed output, respectively. Typically, The value of $N$ is relatively small, i.e., fewer
than 50 samples, which is significantly smaller than the size of the training set in previous fine-tuned methods~\cite{DBLP:conf/emnlp/FengGTDFGS0LJZ20,DBLP:conf/emnlp/0034WJH21}. This setting is referred to as \textit{few-shot in-context learning}. Specially, when the value of $N$ is zero, it is called the \textit{zero-shot in-context learning setting}. Then, ICL concatenates the reconstructed demonstration examples $d_1$ to $d_N$ literally into demonstration  $D=x'_1\parallel y'_1\parallel x'_2\parallel y'_2\parallel ... \parallel x'_N \parallel y'_N$, and further adds the test sample at the end to construct the input prompt $P=D\parallel x'_{test}$, where $\parallel$ denotes the literal concatenation operation. This prompt is finally
fed into the language model for predicting the label $y_{test}$ for test samples. 

Previous studies in NLP 
have shown that the performance of ICL is strongly dependent on the quality of the demonstration. For example, Liu et al.~\cite{DBLP:conf/acl-deelio/LiuSZDCC22} show that selecting demonstration examples with higher similarity or increasing the number of demonstration examples can improve ICL's performance.
The results in~\cite{DBLP:conf/acl/LuBM0S22} show that the order of demonstration examples also has a large impact on the results. 
Following previous studies, we summarize three key factors to consider when designing a demonstration for ICL: the selection, ordering, and number of demonstration examples, as shown in Fig.~\ref {fig:example}.

We would like to further clarify that there are two types of demonstration in ICL: \textbf{task-level} demonstration and \textbf{instance-level} demonstration~\cite{DBLP:conf/naacl/RubinHB22,DBLP:journals/corr/abs-2209-01975}. 
The task-level demonstration uses the same demonstration examples for all test samples and does not take the difference of each test sample into consideration, while the instance-level demonstration selects 
different demonstration examples for different test samples. Although instance-level demonstrations generally perform better than task-level demonstrations, it requires a labeled training set in advance for retrieval. The task-level demonstration is more flexible as it can be used in scenarios where very few data are labeled, or no labeled data are available by selecting few representative data for human labeling~\cite{DBLP:journals/corr/abs-2209-01975}.  In this paper, we investigate both the task-level and instance-level demonstration construction methods for code intelligence tasks.

\section{EXPERIMENTAL EVALUATION}\label{sec:setup}


\subsection{Research Questions}

We design experiments to investigate the impact of the selection, ordering, and number of demonstrations on ICL for code intelligence tasks. Our research aims to answer the following questions:

\begin{enumerate}[label=\bfseries RQ\arabic*:,leftmargin=.5in]
    \item What kind of selection methods are helpful for ICL in code intelligence tasks?
    \item How should demonstration examples be arranged for ICL in code intelligence tasks?
    \item How does the number of demonstration examples in a prompt impact the performance of ICL in code intelligence tasks?
    \item How is the generalizability of our findings? 
\end{enumerate}
In RQ1, we aim at verifying whether selecting similar and diverse demonstration examples is helpful. Besides, we also compare different retrieval methods to analyze the impact of different similarity measurement methods
for ICL.
RQ2 aims at investigating the influence of ordering methods by comparing random ordering with similarity-based ordering.
In RQ3, we want to explore whether increasing the number of examples could bring better performance for ICL.
In RQ4, we evaluate
whether the findings achieved in RQ1-RQ3
are also applicable to different LLMs for verifying the generalizability of the findings.

\subsection{Evaluation tasks}
We conduct experiments on three popular
code intelligence tasks: code summarization, bug fixing, and program synthesis.

\subsubsection{Code Summarization} Code summarization, also known as code comment generation, aims to generate useful comments automatically for a given code snippet~\cite{DBLP:conf/ijcai/HuLXLLJ18}. Recent work mainly formulates it as a sequence-to-sequence neural machine translation (NMT) task and involves pre-trained techniques to achieve better performance~\cite{DBLP:conf/emnlp/0034WJH21,DBLP:conf/icse/GaoZGW23}.

\textbf{Datasets.} To evaluate the performance of code summarization, we use two widely-used datasets: CodeSearchNet (CSN)~\cite{DBLP:journals/corr/abs-1909-09436} and TLCodeSum (TLC)~\cite{DBLP:conf/ijcai/HuLXLLJ18}. CSN is a large-scale source code dataset mined from open-source GitHub repositories. It contains code summarization data in six programming languages, i.e., Java, Go, JavaScript, PHP, Python, and Ruby. The dataset is split into training, validation, and test sets in the proportion of 8:1:1. In this study, considering our time and resource limitation, we use the Java portion of the filtered CSN dataset in CodeBERT~\cite{DBLP:conf/emnlp/FengGTDFGS0LJZ20}, which contains 181,061 samples across the training, validation, and test sets for evaluation. TLC has 87,136 code-comment pairs crawled from 9,732 open-source Java projects on GitHub with at least 20 stars. The code snippets are all at the method level and the comments of corresponding Java methods are considered as code summaries. The portion of training, validation, and test set is also 8:1:1. As reported in previous work, there are duplicated data in the training and test set. Therefore, we follow previous work~\cite{DBLP:conf/icse/ShiWD0H00S22} and remove the duplicated data, and finally get a test set with 6,489 samples.

\textbf{Metrics.} We use three widely-adopted metrics for code summarization evaluation: BLEU-4~\cite{DBLP:conf/acl/PapineniRWZ02}, ROUGE-L~\cite{lin-2004-rouge} and METEOR~\cite{DBLP:conf/acl/BanerjeeL05} for evaluation. These metrics evaluate the similarity between generated summaries and ground-truth summaries and are widely used in code summarization~\cite{DBLP:conf/acl/AhmadCRC20,DBLP:conf/icse/ShiWD0H00S22,DBLP:journals/tosem/GaoGHZNXL23}.

\subsubsection{Bug Fixing} Bug fixing is the task of automatically fixing bugs in the given code snippet. It helps software developers find and fix software errors~\cite{DBLP:conf/icse/TufanoPWBP19,DBLP:journals/tosem/TufanoWBPWP19}.

\textbf{Datasets.} The dataset for bug fixing is B2F which is collected by Tufano et al.~\cite{DBLP:conf/icse/TufanoPWBP19} from bug-fixing commits in GitHub. We use the multi-model version proposed in MODIT~\cite{DBLP:conf/kbse/ChakrabortyR21} for experiments as it contains both the code changes and the fix instruction. 
The model is given both the buggy code and natural language fix guidance to predict the fixed code. We follow their original setting to split the dataset into two parts B2F$_{medium}$ and B2F$_{small}$ based on the length of code tokens (the code length of B2F$_{medium}$ is between 50 and 100 tokens and that of B2F$_{small}$ is below 50 tokens).

\textbf{Metrics.} We follow previous work~\cite{DBLP:conf/sigsoft/ChakrabortyADDR22} and use Exact Match (EM) and BLEU-4 for both datasets.

\begin{table}[t]
 \centering
 \aboverulesep=0ex
\belowrulesep=0ex
\caption{Statistics of the benchmark datasets.}
\label{tab:dataset}
\scalebox{1.05}{
\begin{tabular}{l|lrrr}
\toprule
\textbf{Task} & \textbf{Datasets} & \textbf{Train } & \textbf{Dev} & \textbf{Test}\\ 
\midrule
\multirow{2}{*}{Code Summarization} & {CSN-Java } & 164,923 & 5,183 & 10,955\\
& {TLC} & 69,708 & 8,714 & 6,489\\
\midrule
\multirow{2}{*}{Bug Fixing} & {B2F$_{small}$} & 46,628 & 5,828 & 5,831\\
& {B2F$_{medium}$} & 53,324 & 6,542 & 6,538\\
\midrule
Program Synthesis & {CoNaLa} & 2,389 & - & 500\\ 
\bottomrule
\end{tabular}}
\end{table}

\subsubsection{Program Synthesis} Program synthesis is the task of generating source code based on the given natural language description. It provides practical assistance to developers and enhances their productivity~\cite{DBLP:journals/corr/abs-2107-03374}.

\textbf{Datasets.} For program synthesis, we use the CoNaLa~\cite{DBLP:conf/msr/YinDCVN08} dataset for evaluation. This dataset consists of 2,889 $\langle$intent, code$\rangle$ pairs mined from Stack Overflow in Python. We directly use the original partition of the dataset, which includes 2,389 samples for training and 500 samples for testing.

\textbf{Metrics.} We follow previous work~\cite{DBLP:conf/sigsoft/ChakrabortyADDR22} and evaluate the performance of program synthesis with four metrics including Exact Match (EM), CodeBLEU (CB), Syntax Match (SM), and Dataflow Match (DM).  EM measures whether the code generated by the model is identical to the goal code. CB~\cite{DBLP:journals/corr/abs-2009-10297} is a modified version of BLEU designed specifically for code, which leverages syntax and semantic information such as Abstract Syntax Tree (AST) and data flow to measure the similarity of two code snippets. 
SM and DM are two components that calculate the matching subtrees and data flow edges' proportion, respectively.

\begin{table}[t]
\renewcommand{\arraystretch}{1.25}
 \centering
 \aboverulesep=0ex
\belowrulesep=0ex
\caption{Prompt template for each task. Here text in the form of \underline{\{\#xxx\}} will be filled in actual inputs from the dataset.}
\label{tab:template}
\scalebox{0.9}{
\begin{tabular}{l|l}
\toprule
\textbf{Task} & \textbf{Template}\\ 
\midrule
\multirow{2}{*}{Code Summarization} & Generate comment (summarization) for this code \\
 & [input] \underline{\{\#code\}} [output] \underline{\{\#comment\}} \\
\midrule
\multirow{2}{*}{Bug Fixing} & Fix the bug according to the guidance [input] \\
 & \underline{\{\#buggy code\}} $<$s$>$ \underline{\{\#instruction\}} [output] \underline{\{\#fixed code\}} \\
\midrule
\multirow{2}{*}{Program Synthesis} & Generate code based on the requirement \\
 & [input] \underline{\{\#requirement\}}[output] \underline{\{\#code\}}\\
\bottomrule
\end{tabular}}
\end{table}

\subsection{Implementation}\label{sec:detail}
We utilize the OpenAI Codex (code-davinci-002) API~\cite{DBLP:journals/corr/abs-2107-03374} in our paper for all experiments in the first three RQs. In RQ4, we further use the API of GPT-3.5 (text-davinci-003)~\cite{DBLP:conf/nips/BrownMRSKDNSSAA20} and ChatGPT (gpt-3.5-turbo)~\cite{ChatGPT} for experiments. As for the hyperparameters of the APIs, following the previous work~\cite{DBLP:journals/corr/abs-2210-02875,nashidretrieval}, 
we set the temperature to 0 to get the deterministic output. The frequency\_penalty and presence\_penalty are also set to 0. 
The input length limitation
of Codex, GPT-3.5, and ChatGPT is 8,001, 4,096, and 4,097 tokens, respectively. Hence we cut off the input code of each demonstration example to $\frac{8001}{N+1}$, $\frac{4096}{N+1}$, and $\frac{4097}{N+1}$ tokens, respectively, where ${N}$ represents the number of demonstration examples.  
Empirically, it took approximately 6 hours to evaluate 1,000 examples for Codex.  
To avoid excessive time costs, we randomly sample a small test set (2,000 samples) for each dataset with over 2,000 test samples. We use four
examples in the demonstration
in RQ1 and RQ2, and further discuss the impact of the number of demonstration examples in RQ3. The templates used
in this study are shown in Table~\ref{tab:template}. We also show some examples in our GitHub repository\footnote{https://github.com/gszsectan/ICL/tree/master/prompts}.
We conduct all the experiments on a server with 2 Nvidia RTX 3090 GPUs. The GPUs are used in the dense retrieval process.

\section{Experimental Results}\label{sec:result}

\begin{table*}[t]
\centering
\caption{Experimental results of different demonstration selection methods on  Code Summarization. ``Avg'' and ``CV'' denote the average results and Coefficient of Variation over three different orders, respectively.}\label{tab:RQ1_sum}
\aboverulesep=0ex
\belowrulesep=0ex
\scalebox{1}{
\begin{tabular}{l|cccccc|cccccc}
\toprule 
\multirow{2}{*}{\textbf{Approach}} &\multicolumn{12}{c}{\textbf{Code Summarization}}\\ 
\cmidrule{2-13}
&\multicolumn{6}{c|}{\textbf{CSN}} &\multicolumn{6}{c}{\textbf{TLC}}\\
\cmidrule{2-13}
& \multicolumn{2}{c}{\textbf{BLEU-4}} & \multicolumn{2}{c}{\textbf{ROUGE-L}} & \multicolumn{2}{c|}{\textbf{METEOR}} & \multicolumn{2}{c}{\textbf{BLEU-4}} & \multicolumn{2}{c}{\textbf{ROUGE-L}} & \multicolumn{2}{c}{\textbf{METEOR}}\\
& Avg & CV & Avg & CV & Avg & CV & Avg & CV & Avg & CV & Avg & CV\\
\midrule
& \multicolumn{12}{c}{\textbf{Task-level Demonstration}}\\
\midrule
 {Random}  & 19.64 & 1.44 & 35.46 & 1.88 & 15.30 & 1.54 & 17.29 & 0.71 & 34.28 & 0.61 &12.48 & 0.67 \\
 {KmeansRND}  & \textbf{20.71} & 0.82 & \textbf{38.03} & 0.44 & \textbf{16.34} & 0.83 & \textbf{17.91} & 1.19 & \textbf{35.69} & 1.60 & \textbf{13.48} & 0.91 \\
\midrule
& \multicolumn{12}{c}{\textbf{Instance-level Demonstration}}\\
 \midrule
 {BM-25}  & \textbf{22.35} & 0.46 &38.31 & 0.56 & \textbf{17.01} & 0.78 & \textbf{36.96} & 0.84 & \textbf{51.42} & 0.79 & \textbf{24.22} & 0.99\\
 {SBERT}  & 22.27 & 0.23 & \textbf{38.39} & 0.42 & 16.91 & 0.22 & 36.42 & 0.61 &50.47 & 0.40 &23.86 & 0.68\\
 {UniXcoder}  & 22.11 & 0.61 &38.23 & 0.53 & 16.81 & 0.23 & 36.77 & 0.52 &51.11 & 0.29 &24.08 & 0.79\\
 {CoCoSoDa}  & 21.92 & 0.46 &37.85 & 0.22 & 16.78 & 0.24 & 36.91 & 0.69 &50.69 & 0.53 &24.08 & 0.39\\
 {Oracle  (BM-25)}  & 27.69 & 0.43 &46.17 & 0.14 & 20.26 & 0.22 & 43.16 & 0.15 &59.17 & 0.09 &28.09 & 0.16\\
\bottomrule
\end{tabular}
}
\end{table*}

\subsection{RQ1: Demonstration Selection}

\subsubsection{Experimental design}
We first explore the impact of demonstration selection methods on ICL for code-related tasks. To provide a comprehensive study, we adopt 
different kinds of demonstration selection methods for the three code intelligence tasks. 

For task-level demonstration, we need to select a group of demonstration examples for the whole test set, as illustrated in Section~\ref{sec:back_icl}. To explore the influence of
different in-context demonstration examples
on the performance of ICL, we randomly select three groups of demonstration examples from the training set,
and evaluate their performance on different tasks, denoted as \textit{Random}. Besides, we further investigate whether improving the diversity of demonstration examples is beneficial to ICL. We
select the demonstration examples by first dividing the whole samples into $N$ clusters and then randomly selecting one sample from each 
cluster, namely \textit{KmeansRND}. Specifically, we use UniXcoder~\cite{DBLP:conf/acl/GuoLDW0022} for vectorization and use the K-means++ algorithm~\cite{DBLP:conf/soda/ArthurV07} for clustering, where $K$ is set to $N$ that represents the number of demonstration example. 
Similar to \textit{Random},
we also investigate the performance of different groups of examples for \textit{KmeansRND} and conduct the selection process three times, resulting in three groups of demonstration examples.

For instance-level demonstration, we need to select examples for each test sample, as illustrated in Section~\ref{sec:back_icl}. 
Following~\cite{DBLP:conf/acl-deelio/LiuSZDCC22}, we formulate the selection process
as a retrieval problem and compare the performance of different retrieval-based methods including:

\begin{table}[t]
\centering
\caption{Experimental results of different demonstration selection methods on Bug Fixing.
}\label{tab:RQ1_bug}
\aboverulesep=0ex
\belowrulesep=0ex
\scalebox{0.89}{
\begin{tabular}{l|ccrr|ccrc}
\toprule
\multirow{4}{*}{\textbf{Approach}} & \multicolumn{8}{c}{\textbf{Bug Fixing}} \\ 
\cmidrule{2-9}
& \multicolumn{4}{c|}{\textbf{B2F$_{medium}$}} & \multicolumn{4}{c}{\textbf{B2F$_{small}$}}\\
\cmidrule{2-9}
& \multicolumn{2}{c}{\textbf{BLEU-4}} & \multicolumn{2}{c|}{\textbf{EM}} & \multicolumn{2}{c}{\textbf{BLEU-4}} & \multicolumn{2}{c}{\textbf{EM}} \\
& Avg & CV & Avg & CV & Avg & CV & Avg & CV\\
\midrule
& \multicolumn{8}{c}{\textbf{Task-level Demonstration}}\\
\midrule
 {Random}  & \textbf{86.96} & 0.16 & 7.26 & 16.18 & 71.18 & 0.56 & 9.95 & 6.33 \\
 {KmeansRND} & 86.91 & 0.17 & \textbf{9.03} & 5.45  & \textbf{72.89} & 1.36 & \textbf{10.37} & 3.86 \\
\midrule
& \multicolumn{8}{c}{\textbf{Instance-level Demonstration}}\\
 \midrule
 {BM-25}  & \textbf{88.05} & 0.09 & \textbf{21.85} & 1.78 & \textbf{77.54} & 0.13 & \textbf{30.45} & 0.96 \\
 {SBERT}  & 87.98 & 0.06 & 19.00 & 2.88  & 76.26 & 0.16 & 26.15 & 0.87 \\
 {UniXcoder}  & 87.87 & 0.09 & 19.14 & 2.00 & 77.52 & 0.07 & 29.93 & 0.51 \\
 {CoCoSoDa}  & 87.73 & 0.07 & 19.23 & 0.74 & 76.45 & 0.07 & 27.40 & 1.04 \\
\bottomrule
\end{tabular}
}
\end{table}

\begin{enumerate}
\item \textbf{BM-25}: BM-25 is a classic sparse retrieval method in the information retrieval field. It has also been widely used in many 
code intelligence models~\cite{DBLP:conf/icse/ZhangW00020,DBLP:conf/kbse/WeiLLXJ20}.

\item \textbf{SBERT}: SBERT~\cite{DBLP:conf/emnlp/ReimersG19} is a popular sentence modeling method and has been widely used in text retrieval~\cite{DBLP:journals/tacl/LuanETC21,DBLP:conf/emnlp/ReimersG19}.  Specifically, in this paper, we use the version 
that is further trained on the code-related dataset to obtain code and text representations~\cite{flaxcode}.

\item \textbf{UniXcoder}: UniXcoder~\cite{DBLP:conf/acl/GuoLDW0022} is a unified cross-modal pre-trained model that is pre-trained with three sequence modeling tasks and two contrastive learning-based tasks. It shows promising performance on zero-shot code-to-code search.
\item \textbf{CoCoSoDa}: CoCoSoDa~\cite{shi2022enhancing} is a state-of-the-art code search model that utilizes contrastive learning for code and text representation learning.
\end{enumerate}

For BM-25, we implement 
with the gensim package~\cite{Gensim}
by retrieving samples with the highest similarity from the training set. 
For dense retrieval methods, we directly use these pre-trained models in the replication packages released by the authors without further tuning. Based on the code/text representations output by the pre-trained models, we select the training samples presenting the highest cosine similarities with the test sample.
We also follow the previous work~\cite{DBLP:conf/naacl/RubinHB22} and create a method called \textit{Oracle},
which 
selects
demonstration examples by calculating the similarity between
the output of the test sample and
the output of all training set examples. The Oracle method is usually regarded as an upper bound of the performance, considering that the output of the test sample is not available in practice.
The retrieval process in Oracle is implemented by BM-25, since BM-25
shows the best performance compared with other dense retrieval methods as shown in Table~\ref{tab:RQ1_sum}-\ref{tab:RQ1_gen}.  

To avoid the influence of different orders of demonstration examples, we run each experiment three times with different 
orders and report the average results on each metric. Besides, we further evaluate the sensitivity of each method to different orders by Coefficient of Variation (CV)~\cite{Brown1998}. The CV
is calculated by $\sigma / \mu$,
where $\sigma$ is the standard deviation and $\mu$ is the mean. A lower CV indicates smaller data variation.
It takes the magnitude of data into account and has been widely used to measure the data dispersion in many fields such as economics and software engineering~\cite{Brown1998,DBLP:conf/icse/WeiHH0022}. 


\subsubsection{Analysis} 
\begin{figure}
    \centering
    \begin{subfigure}[b]{0.167\textwidth}
      \centering
      \includegraphics[width=1\textwidth]{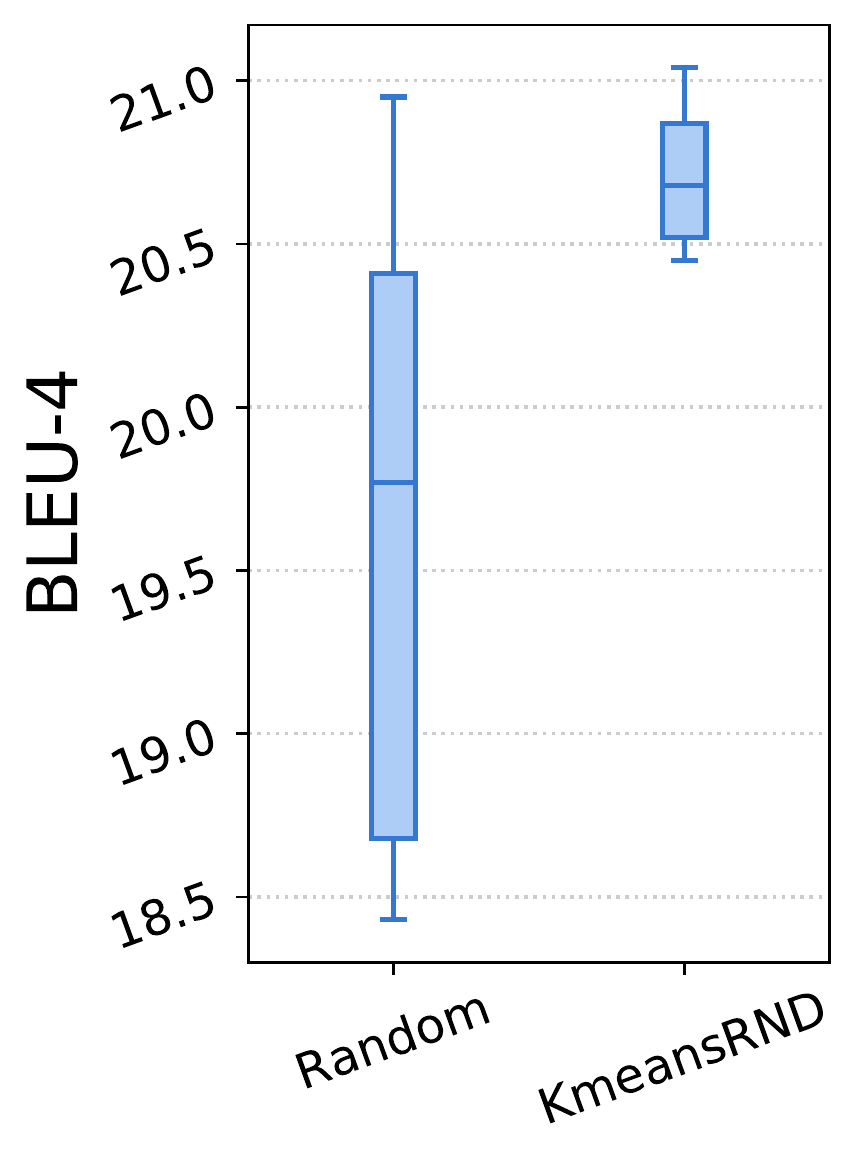}
      \caption{CSN.}
    \end{subfigure}
    \hspace{.2in}
    \begin{subfigure}[b]{0.1495\textwidth}
      \centering
      \includegraphics[width=1\textwidth]{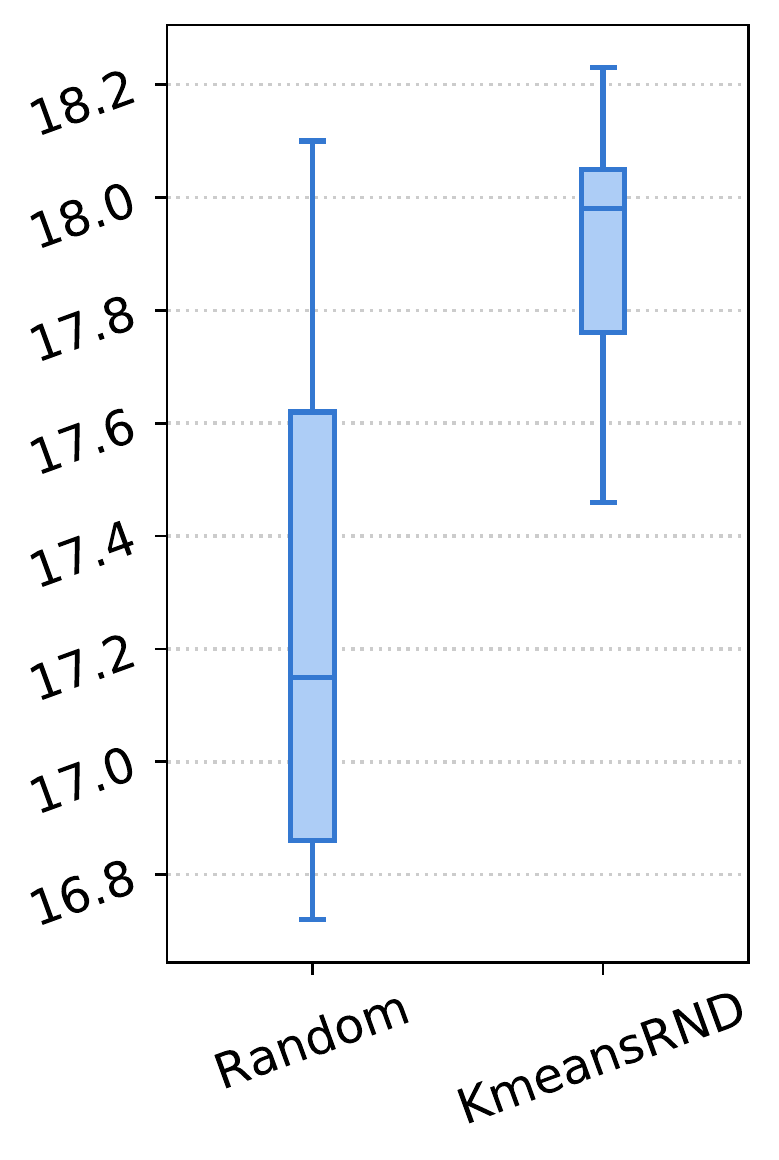}
      \caption{TLC.}
    \end{subfigure}
    \\
    \centering
    \begin{subfigure}[b]{0.167\textwidth}
      \centering
      \includegraphics[width=1\textwidth]{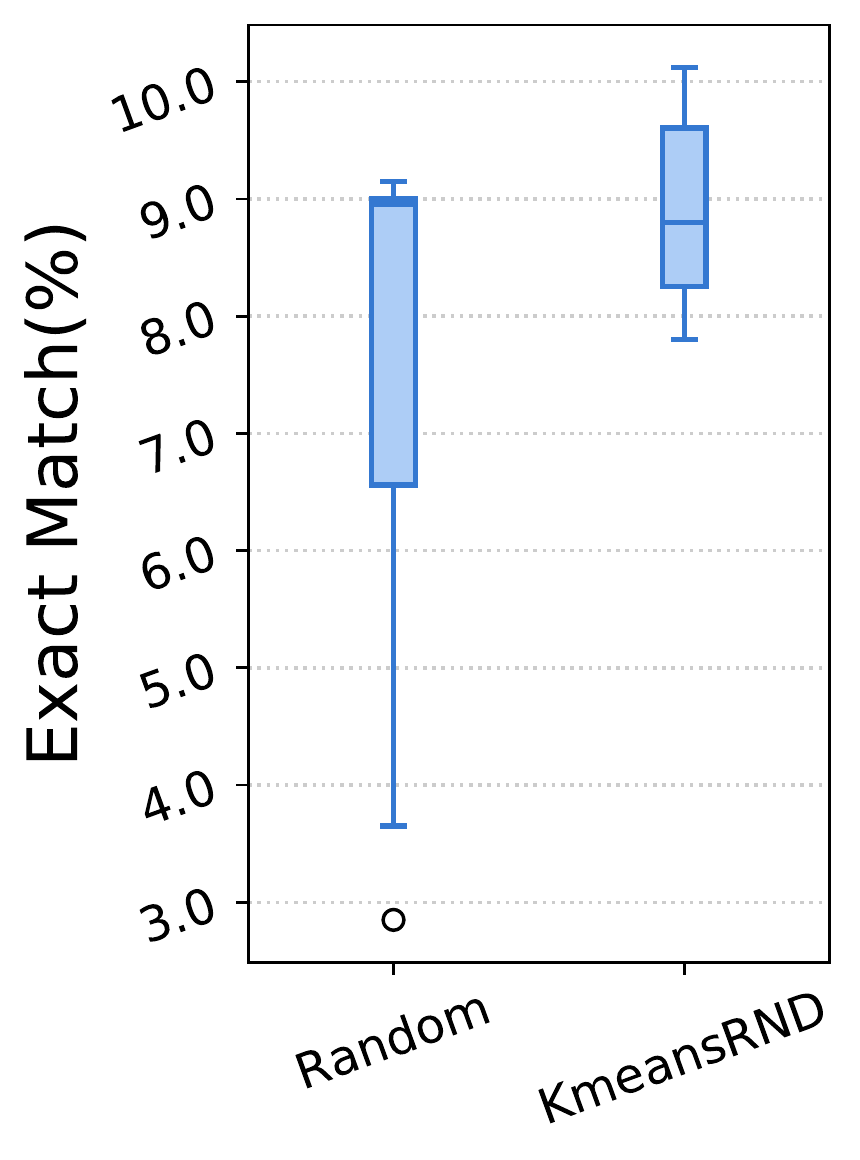}
      \caption{B2F$_{medium}$.}
    \end{subfigure}
    \begin{subfigure}[b]{0.1495\textwidth}
      \centering
      \includegraphics[width=1\textwidth]{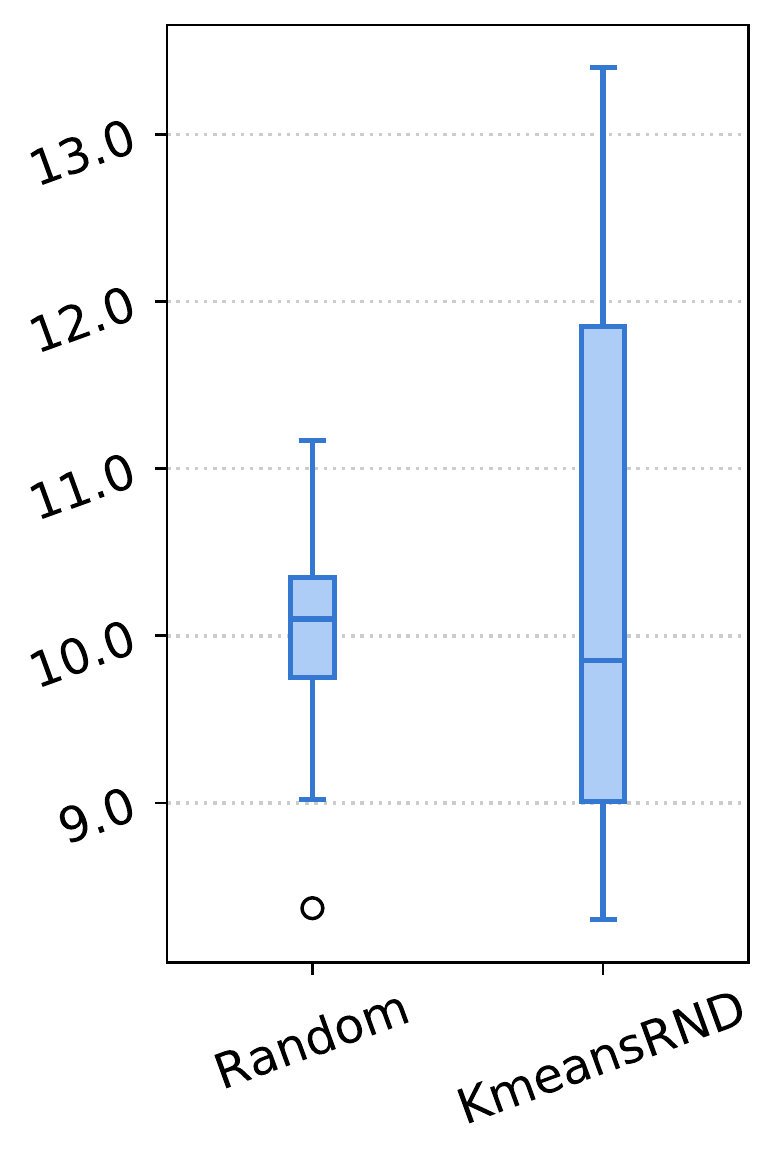}
      \caption{B2F$_{small}$.}
    \end{subfigure}
    \begin{subfigure}[b]{0.1495\textwidth}
      \centering
      \includegraphics[width=1\textwidth]{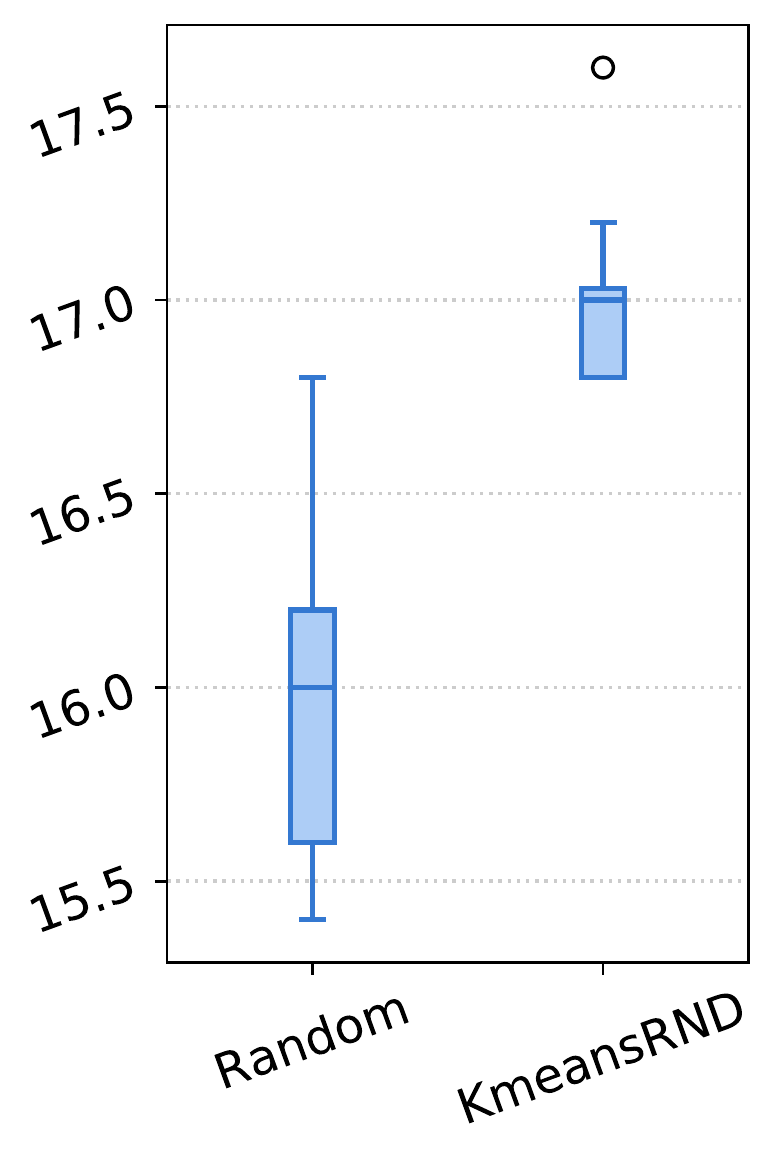}
      \caption{CoNaLa.}
    \end{subfigure}
    \caption{Comparison of the performance distribution of Random and KmeansRND regarding different groups of examples on three tasks.}
    \label{fig:variance}
\end{figure}

We present the experimental results in Table~\ref{tab:RQ1_sum}-\ref{tab:RQ1_gen}. For each metric, we report the average results over three random orders and CV which measures their sensitivity to different orders. 
In Fig.~\ref{fig:variance}, we show the distribution of results with different groups of examples for Random and KmeansRND.

\textbf{Diversity of examples is beneficial for task-level demonstration.} As can be seen in Table~\ref{tab:RQ1_sum}-\ref{tab:RQ1_gen} and Fig.~\ref{fig:variance}, by comparing the results on Random and KmeansRND, we can find that in most cases improving the 
diversity of task-level demonstrations can not only improve the average performance of ICL but also reduce the fluctuation brought by different groups of examples.
For example, as shown in Table~\ref{tab:RQ1_sum}, comparing the results of code summarization on CSN, the average improvements of KmeansRND over Random are 5.45\%, 7.25\%, and 6.80\% with respect to BLEU-4, ROUGE-L, and METEOR, respectively. 
Besides, we can also find that the performance of different in-context
demonstration examples of Random varies
a lot, and improving the diversity of selected examples can reduce this variation in general. For example, as shown in Fig.~\ref{fig:variance} (a), the gap between the best and worst BLEU-4 score of Random is about 2.5 while that of KmeansRND is only about 0.6. 
This indicates that improving the diversity of selected demonstration examples is beneficial for building task-level demonstration. 

\begin{tcolorbox}
\textbf{Finding 1:} 
Diversity of examples is helpful for the demonstration selection of ICL. It can help improve overall performance and lead to a more stable prediction regarding different groups of examples.
\end{tcolorbox}

\textbf{BM-25 is a simple and effective method for instance-level demonstration.} By comparing the results of different instance-level demonstration methods, we can find that the simple BM-25 method can achieve comparable or even better performance than other dense retrieval methods on demonstration selection in ICL. For example, the average EM of BM-25 on Program Synthesis is 18.53, which outperforms three strong dense retrieval methods SBERT, UniXcoder, and CoCoSoDa by 14.88\%, 15.81\%, and 13.68\%, respectively. {This result indicates that BM-25 serves as an effective
baseline approach and could be taken into account in future studies of demonstration selection for code intelligence tasks.}



\begin{table}[t]
\centering
\caption{Experimental results of different demonstration selection methods on Program Synthesis. 
}\label{tab:RQ1_gen}
\aboverulesep=0ex
\belowrulesep=0ex
\scalebox{0.9}{
\begin{tabular}{l|cccccccc}
\toprule
\multirow{3}{*}{\textbf{Approach}} & \multicolumn{8}{c}{\textbf{Program Synthesis}} \\
\cmidrule{2-9}
& \multicolumn{2}{c}{CB} & \multicolumn{2}{c}{SM} & \multicolumn{2}{c}{DM} & \multicolumn{2}{c}{EM} \\
& Avg & CV & Avg & CV & Avg & CV & Avg & CV\\
\midrule
& \multicolumn{8}{c}{\textbf{Task-level Demonstration}}\\
\midrule
 {Random}  & \textbf{28.36} & 1.30 & 44.37 & 0.83 & \textbf{39.70} & 1.33 & 16.00 & 1.60\\
 {KmeansRND}  & 28.03 & 1.47  & \textbf{44.41} & 0.54  & 37.31 & 1.54  & \textbf{17.03} & 1.06  \\
\midrule
& \multicolumn{8}{c}{\textbf{Instance-level Demonstration}}\\
 \midrule
 {BM-25}  &  \textbf{30.37} & 0.91 & \textbf{46.22} & 0.84 & 40.75 & 1.06  & \textbf{18.53} & 0.50  \\
 {SBERT}   & 29.08 & 0.70 &44.91& 0.31 & 39.81 & 3.01 & 16.13 & 2.54 \\
 {UniXcoder}  & 28.96 & 0.50 & 43.93 & 0.67 & 37.96 & 1.12 & 16.00 & 3.53 \\
 {CoCoSoDa}  & 29.42 & 0.82 & 44.62 & 0.70 & \textbf{40.91} & 1.12 & 16.30 & 0.86 \\
\bottomrule
\end{tabular}
}
\end{table}

\begin{tcolorbox}
\textbf{Finding 2:}
The retrieval methods for demonstration selection can impact the performance of ICL, among which BM-25 is a simple and effective method.

\end{tcolorbox}

\begin{table*}[t]
    \centering
    \caption{Experimental results of different demonstration ordering methods.}
\aboverulesep=0ex
\belowrulesep=0ex
 \scalebox{1}{
    \begin{tabular}{cc|rrr|rr|rrrr}
    \toprule
    \multicolumn{2}{c|}{{\multirow{2}{*}{\textbf{Approach}}}} & \multicolumn{3}{c|}{Code Summarization (CSN)} & \multicolumn{2}{c|}{Bug Fix (B2F$_{small}$)} & \multicolumn{4}{c}{Program Synthesis (CoNaLa)} \\
    \cmidrule{3-11} 
     &  &BLEU-4 & ROUGE-L & METEOR  & BLEU-4 & EM & \multicolumn{1}{c}{CB} & \multicolumn{1}{c}{SM} & DM & EM  \\
    \midrule
    \multirow{3}{*}{Random} & \multicolumn{1}{|c|}{Random} & 20.46 & 36.71 & 16.17 & \textbf{72.40} & 9.52  & 27.72 & 44.46 & 37.53 & 15.53 \\
    \multicolumn{1}{c|}{}& \multicolumn{1}{c|}{Similarity} & \textbf{21.04} & \textbf{37.86} & \textbf{16.26} & 72.02 & \textbf{9.93} & \textbf{28.47} & \textbf{44.87} & 37.79 & \textbf{16.00} \\
    \multicolumn{1}{c|}{}& \multicolumn{1}{c|}{Reverse Similarity} & 19.78 & 33.71 & 15.64 & 71.44 & 9.02  & 27.62 & 44.48 & \textbf{37.96}& 15.20 \\
    \midrule
    \multirow{3}{*}{KmeansRND} & \multicolumn{1}{|c|}{Random}  & 20.67 & \textbf{37.64} & 15.97 & 72.29 & 8.60 & 26.64 & 42.97 & \textbf{37.24} & \textbf{16.87}\\
    \multicolumn{1}{c|}{}& \multicolumn{1}{c|}{Similarity} & \textbf{20.69} & 37.62 & 16.05 & \textbf{72.90} & \textbf{10.15} &\textbf{27.20} &42.97 &36.93& 16.40  \\
    \multicolumn{1}{c|}{}& \multicolumn{1}{c|}{Reverse Similarity } & 20.55 & 37.43 & \textbf{16.20} & 72.05 & 9.78 &27.09 & \textbf{43.74} &37.19 & 16.60\\
    \midrule
    \multirow{3}{*}{BM-25} & \multicolumn{1}{|c|}{Random} & \textbf{22.35} & \textbf{38.31} & \textbf{17.01} & 77.54 & 30.45 & 30.37 & 46.22 & 40.75 & \textbf{18.53} \\
    \multicolumn{1}{c|}{}& \multicolumn{1}{c|}{Similarity} & 22.23 & 38.12 & \textbf{17.01} & \textbf{77.76} & \textbf{30.95} &\textbf{30.83} & \textbf{46.41} & \textbf{41.33} & 17.60 \\
    \multicolumn{1}{c|}{}& \multicolumn{1}{c|}{Reverse Similarity } & 22.13 & 38.26 & 16.91 & 77.60 & 29.80 & 30.01 & 45.72 & 39.60 & 18.20 \\
    \bottomrule
    \end{tabular}
    \label{tab:order}}
\end{table*}

\textbf{Instance-level demonstration outperforms task-level demonstration greatly.
} As shown in Table~\ref{tab:RQ1_sum}-\ref{tab:RQ1_gen}, we can find that instance-level demonstration can achieve much better performance in all tasks. 
Specifically, the instance-level selection methods improve the best task-level demonstration's exact match results by at least 141.97\% and 193.64\% on B2F$_{medium}$ and B2F$_{small}$, respectively. These results indicate that selecting similar demonstration examples specifically for each test sample can benefit ICL in code intelligence tasks a lot. 


\textbf{The task-level demonstration is more sensitive to the order than the instance-level demonstration.}
By comparing the CV of task-level demonstration and instance-level demonstration, we can find that the performance of instance-level demonstration is generally more stable than task-level demonstration regarding different example orders. Specifically, as shown in Table~\ref{tab:RQ1_bug}, the CV of BLEU-4 of task-level demonstration KmeansRND to the order is 0.17 and 1.36 on two bug fixing datasets, which is 
much larger than that of instance-level demonstration methods (e.g., 0.09 and 0.13 for BM-25, respectively).
{This indicates that selecting examples by similarity is more robust to the changes in the demonstration order and we should carefully arrange the order of demonstration examples  when using task-level demonstration}

\begin{tcolorbox}
\textbf{Finding 3:} 
Compared with task-level demonstration, instance-level demonstrations can achieve much better performance and are generally more robust to the changes in the demonstration order.
\end{tcolorbox}

Apart from the above, we can also observe in Table~\ref{tab:RQ1_sum} that the best demonstration selection method BM-25 still has a large gap with the Oracle. 
This indicates that these retrieval methods may fail to select semantic similar examples and there exists a large space for further improvement concerning the demonstration selection method for code intelligence tasks.

\begin{figure*}[ht]
    \centering
      \hfill
      \begin{subfigure}[b]{0.31\textwidth}
        \centering
         \includegraphics[width=1\textwidth]{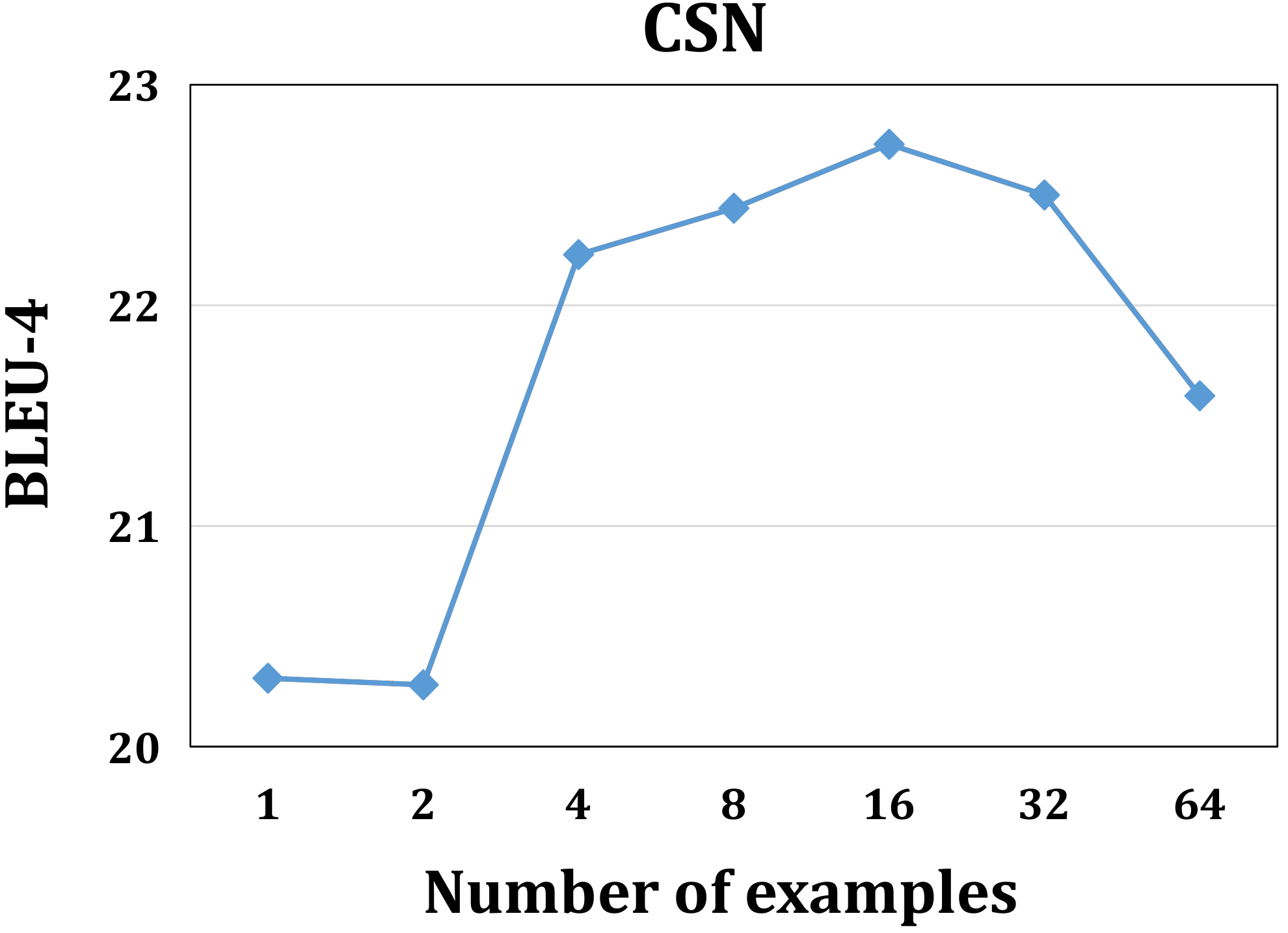}
        \caption{Code Summarization.}
      \end{subfigure}
      \hfill
      \begin{subfigure}[b]{0.31\textwidth}
        \centering
        \includegraphics[width=1\textwidth]{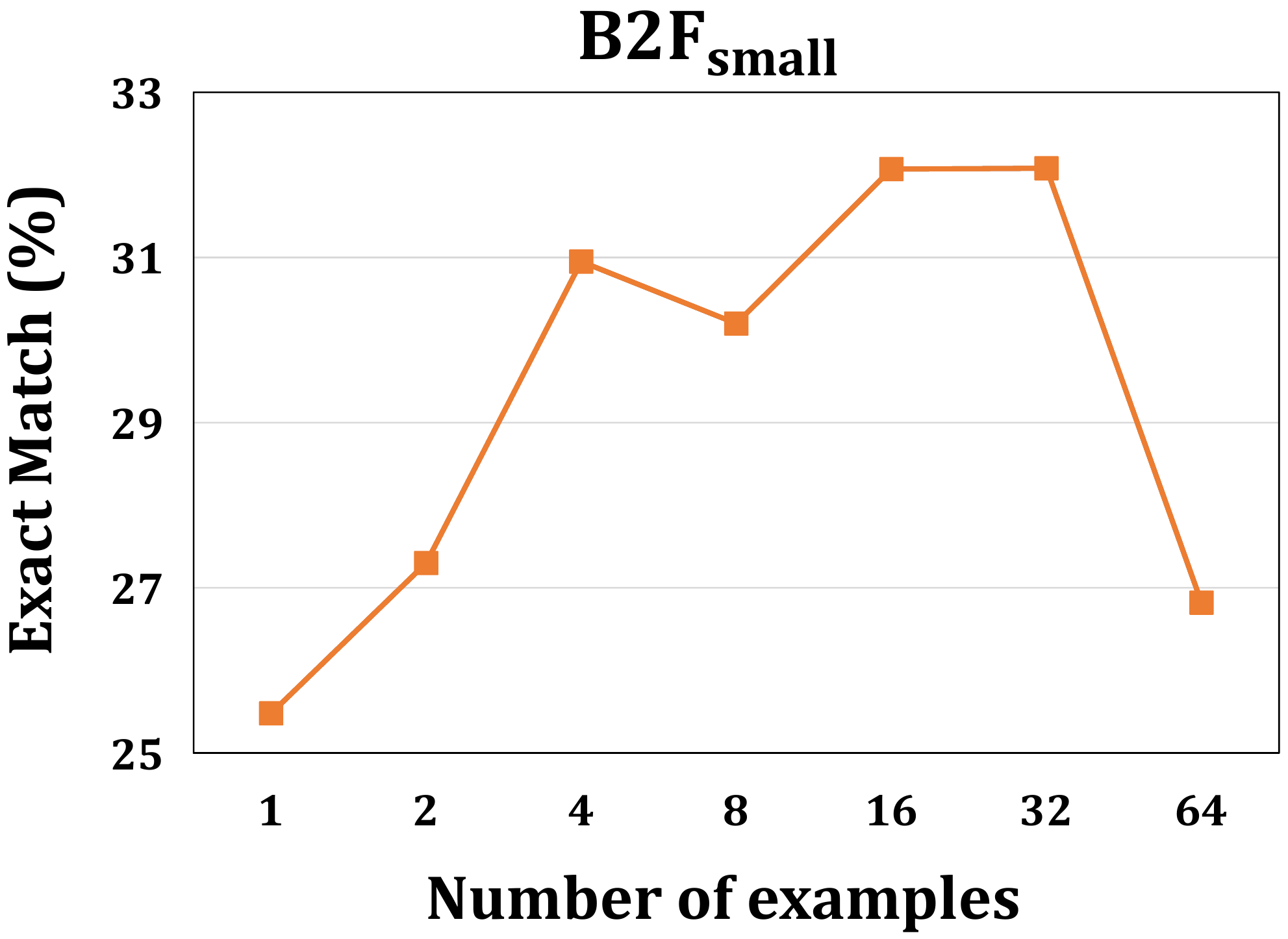}
        \caption{Bug Fixing.}
      \end{subfigure}
      \hfill
      \begin{subfigure}[b]{0.31\textwidth}
        \centering
         \includegraphics[width=1\textwidth]{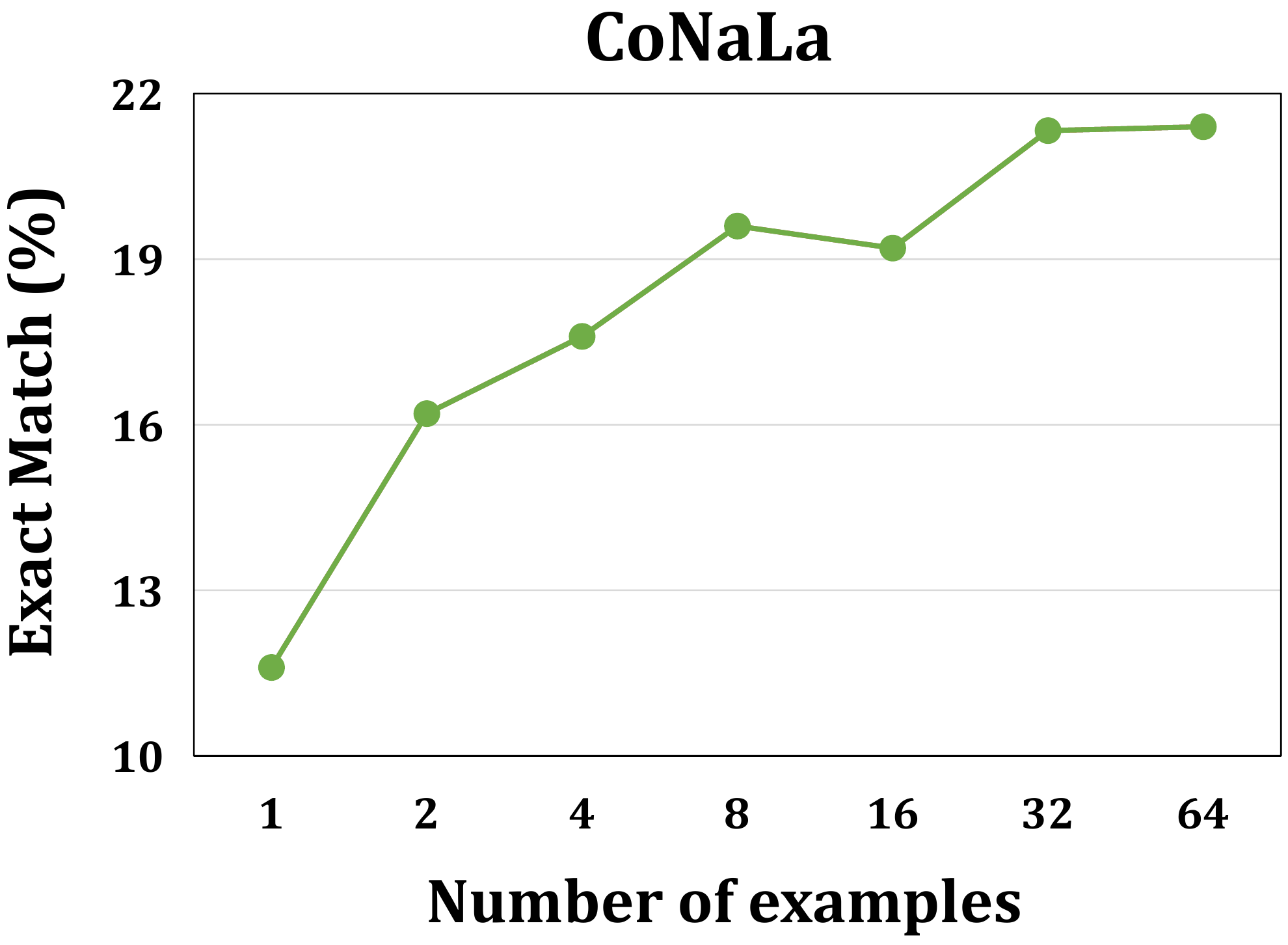}
        \caption{Program Synthesis.}
      \end{subfigure}
    \caption{Experimental results of ICL with different number of demonstration examples.} 
	\label{fig:number}
\end{figure*}

\subsection{RQ2: Demonstration Order}

\subsubsection{Experimental setup} 
In RQ1, we have found that the order of demonstration examples impacts
the performance of ICL on code intelligence tasks, especially on task-level demonstration. Therefore, in this section, we explore how to better arrange the demonstration examples in ICL.
Inspired by the finding
that the task-level demonstration is more sensitive
to the example order than the instance-level demonstration, we suppose that the order of similarities between each
demonstration example and test sample plays an important role
in ICL. 

To verify this, in this RQ, we compare random order with two basic
ordering methods, i.e.,  \textit{Similarity} and \textit{Reverse Similarity}. In the Similarity method, we compare the similarity of each example with test sample 
and the example with a higher similarity will be placed closer to the test sample. On the contrary, for the Reverse Similarity method, the demonstration examples will be placed in descending order according to their similarity to the test sample. 
We experiment with three demonstration selection methods here. As illustrated in RQ1, since the order arrangement is important for task-level demonstration, we use both the Random and KmeansRND for experiments. As for instance-level demonstration, we conduct experiments on BM-25, 
since it shows the best performance among
all the instance-level demonstration selection methods.

\subsubsection{Analysis}
From the results in Table~\ref{tab:order}, we can find that placing the demonstration examples in ascending order based on their similarity to the test sample performs generally better than the reverse. Specifically, \textit{Similarity} consistently outperforms \textit{Reverse Similarity} on code summarization and bug fixing by at least 0.45\% and 0.21\% with respect to BLEU-4 and EM, respectively. By further comparing all the results together, we can observe that similarity achieves the best performance in most cases. Specifically, it achieves the best performance in 62.96\% (17/27) metrics and tasks. However, we can also observe that there are some cases in which both \textit{Similarity} and \textit{Reverse Similarity} perform worse than the average results of using random order, indicating that more complex demonstration ordering methods can be explored by the future work. 

\begin{tcolorbox}
\textbf{Finding 4:} 
The different orders of demonstration examples can impact the performance of ICL. Arranging the demonstration examples based on their similarity to the test sample in ascending order can achieve relatively better results in most cases.
\end{tcolorbox}

\subsection{RQ3: The Number of Demonstration Examples}

\subsubsection{Experimental setup}
In this section, we investigate whether the increase in the number of examples will improve the performance of ICL on code intelligence tasks. We vary the number of demonstration examples from 1 to 64. We use BM-25 and \textit{Similarity}
as demonstration selection and demonstration ordering methods, respectively, based on the above findings.

\subsubsection{Analysis} 
As shown in Fig.~\ref{fig:number}, we can find that the performance of ICL on all the tasks increases with the number of demonstration examples at first. However, when the number of examples is above 16, the results on different tasks show different trends. For example, for bug fixing, the performance achieves the peak when the number of demonstration examples is 32 and suffers from a significant drop when further increasing the number
to 64. As for program synthesis, the performance keeps increasing and tends to be stable when the number exceeds 32. We believe 
that the different trends are
caused by the \textit{truncation problem}~\cite{DBLP:conf/acl/DaiYYCLS19,bulatov2022recurrent}.
As illustrated in Section~\ref{sec:detail}, when increasing the number of examples, the length of the whole demonstration will increase and the examples might be cut off to avoid exceeding the length limitation of LLMs. Specifically, for the B2F$_{small}$ dataset, all the examples are complete without cutting off when the number of examples is below 32. However, when the number
becomes 32, 2.33\% demonstration examples are cut off.
When further increasing the number to 64, the truncation problem happens on over 80\% examples and 44.32\% characters in those examples are discarded, resulting in a dramatic performance degradation. 
Since the length of samples in CSN and B2F$_{small}$ datasets is 
much larger than that of the CoNaLa dataset, i.e., 557, 492, 101 characters per sample for CSN, B2F$_{small}$, and CoNaLa, respectively, the truncation problem does not appear on program synthesis even though the number grows to 64. Therefore, balancing the number of examples and the ensuing truncation problem is important for ICL.

Since the code is generally much longer than natural language~\cite{DBLP:journals/corr/abs-1909-09436}, the truncation problem is easier to appear in code intelligence tasks. Besides, more examples will also lead to a larger cost of using external API and the inference time~\cite{pricing}. 
A smaller number of examples may be more appropriate for code intelligence tasks.
From the results (Fig.~\ref{fig:number}), we can also find that the performance with four demonstration examples is good enough, achieving
96.48\%, 97.80\%, and 94.80\% of the best performance on the three tasks with respect to EM, BLEU-4, and CodeBLEU, respectively. 
Therefore, considering the above trade-off, 
using four 
examples in the demonstration
is a good choice for code intelligence tasks. 

\begin{table}[t]
\centering
\caption{Experiments of generalization of findings on GPT3.5 and ChatGPT. 
}\label{tab:generalization}
\aboverulesep=0ex
\belowrulesep=0ex
\scalebox{1.15}{
\begin{tabular}{l|l|cccc}
\toprule
\multicolumn{2}{c|}{{\textbf{Approach}}} & \multicolumn{2}{c}{CB} & \multicolumn{2}{c}{EM} \\
\midrule
\multicolumn{1}{l|}{Selection} & \multicolumn{1}{l|}{Order} & Avg & CV & Avg & CV\\
\midrule
\multicolumn{1}{c}{}& & \multicolumn{4}{c}{\textbf{GPT-3.5}}\\
\midrule
Random & {Random}  & 26.60 & 3.01 & 12.32 & 4.73 \\
KmeansRND  & {Random}  & 28.26 & 1.93  & 13.60 & 1.65 \\
UniXcoder  & {Random}  & 30.06 & 0.53 & 13.73 & 1.13 \\
BM-25 & {Random}  & \textbf{30.81} & 1.05  & 14.40 & 1.81 \\
BM-25 & {Similarity}  & 30.69 & 0.00  & \textbf{15.20} &  0.00 \\
\midrule
\multicolumn{1}{c}{}& & \multicolumn{4}{c}{\textbf{ChatGPT}}\\
 \midrule
Random  & {Random}  & 28.17 & 1.98 & 11.88 & 4.24 \\
KmeansRND  & {Random}  & 28.25 & 2.31  & 12.92 & 1.78 \\
UniXcoder  & {Random}  & 29.33 & 1.85 & \textbf{14.32} & 2.87 \\
BM-25  & {Random}  & 28.95 & 5.75  & 13.47 & 1.82 \\
BM-25  & {Similarity}  & \textbf{30.03} & 0.00 & 14.20 & 0.00  \\
\bottomrule
\end{tabular}
}
\end{table}

\begin{table*}[t]
    \centering
    \caption{Comparison of different demonstration construction methods on three LLMs.}
\aboverulesep=0ex
\belowrulesep=0ex
 \scalebox{1}{
    \begin{tabular}{cc|rrr|rr|rrrr}
    \toprule
    \multicolumn{2}{c|}{{\multirow{2}{*}{\textbf{Approach}}}} & \multicolumn{3}{c|}{Code Summarization (CSN)} & \multicolumn{2}{c|}{Bug Fix (B2F$_{small}$)} & \multicolumn{4}{c}{Program Synthesis (CoNaLa)} \\
    \cmidrule{3-11} 
     &  &BLEU-4 & ROUGE-L & METEOR  & BLEU-4 & EM & \multicolumn{1}{c}{CB} & \multicolumn{1}{c}{SM} & DM & EM  \\
    \midrule
    \multirow{3}{*}{Codex} & \multicolumn{1}{|c|}{Zero-shot} & 1.82 & 4.27 & 4.19 & 34.65 & 1.43 & 8.71 & 9.26 & 23.81 & 0.20 \\
    \multicolumn{1}{c|}{}& \multicolumn{1}{c|}{Baseline demonstration}  & 17.37 & 32.04 & 13.43 & 69.07 & 9.70 & 27.54 & 44.56 & 37.07 & 14.20 \\
    \multicolumn{1}{c|}{}& \multicolumn{1}{c|}{Carefully-designed demonstration} & \textbf{22.73} & \textbf{39.52} & \textbf{17.35} & \textbf{77.54}  & \textbf{32.25}  & \textbf{32.07} & \textbf{48.03} & \textbf{42.88}  & \textbf{21.40} \\
    \midrule
    \multirow{3}{*}{GPT-3.5} & \multicolumn{1}{|c|}{Zero-shot} & 6.34 & 15.05 & 14.08 & 2.81 & 0.15 & 0.06& 0.26 & 0.00  & 0.20 \\
    \multicolumn{1}{c|}{}& \multicolumn{1}{c|}{Baseline demonstration}  & 14.55 & 21.53 & 13.81 & 62.87 & 9.15 & 26.36 & 36.94 & 41.67 & 10.00 \\
    \multicolumn{1}{c|}{}& \multicolumn{1}{c|}{Carefully-designed demonstration} & \textbf{15.99} & \textbf{26.78} & \textbf{16.70} & \textbf{71.70} & \textbf{25.25} & \textbf{30.69}  & \textbf{43.95} & \textbf{44.78} & \textbf{15.20} \\
    \midrule
    \multirow{3}{*}{ChatGPT} &\multicolumn{1}{|c|}{Zero-shot} & 3.63 & 11.40 & 13.16 & 2.32 & 0.05 & 25.70 & 37.64 & 54.44 & 3.40 \\
    \multicolumn{1}{c|}{}& \multicolumn{1}{c|}{Baseline demonstration}  & 10.76 & 20.02 & 14.83 & 41.57 & 4.60 & 27.62 & 41.83 & 46.85 & 9.40 \\
    \multicolumn{1}{c|}{}& \multicolumn{1}{c|}{Carefully-designed demonstration} & \textbf{11.90} & \textbf{23.31} & \textbf{16.93} & \textbf{53.92} & \textbf{18.15} & \textbf{30.03} & \textbf{45.04} & \textbf{44.26} & \textbf{14.20} \\
    \bottomrule
    \end{tabular}
    \label{tab:comparison}}
\end{table*}

\begin{tcolorbox}
\textbf{Finding 5:}
More demonstration examples in the prompt will not always lead to better performance considering the truncation problem. 
To save costs, it is suggested that four examples are used in the demonstration.
\end{tcolorbox}

\begin{figure}
    \centering
    \begin{subfigure}[b]{0.2125\textwidth}
      \centering
      \includegraphics[width=1\textwidth]{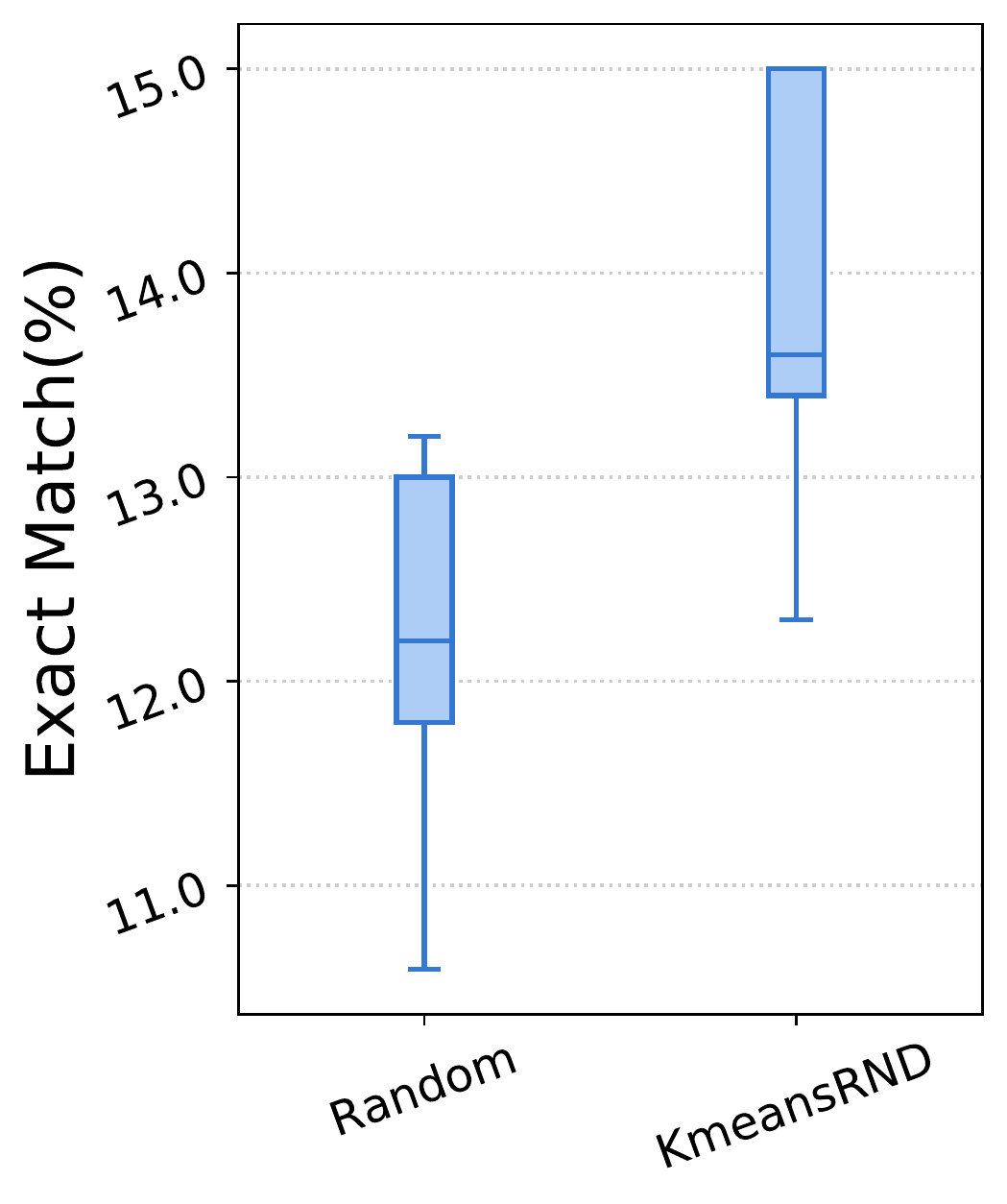}
      \caption{GPT-3.5.}
    \end{subfigure}
    \hspace{.1in}
    \begin{subfigure}[b]{0.1955\textwidth}
      \centering
      \includegraphics[width=1\textwidth]{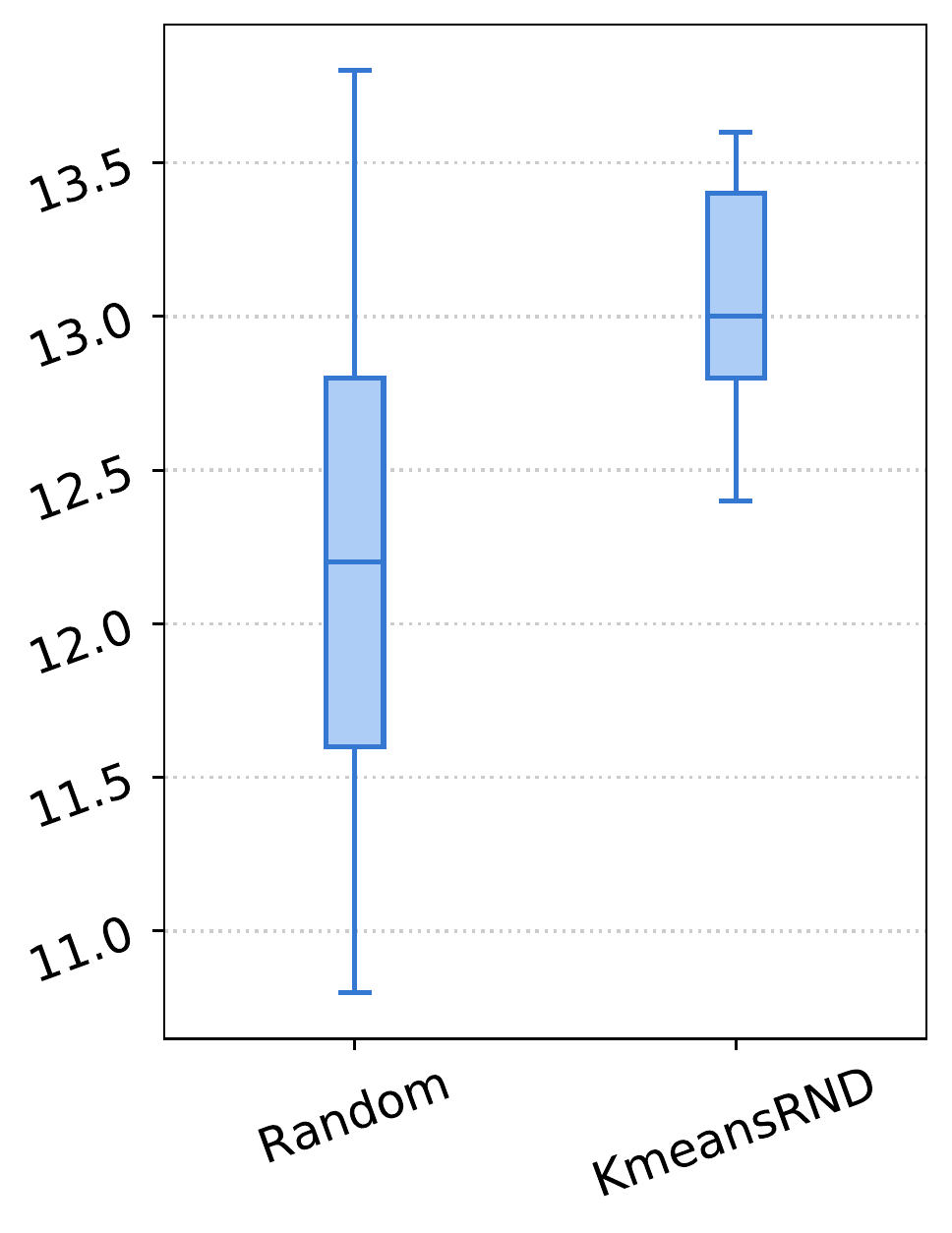}
      \caption{ChatGPT.}
      \end{subfigure}
    \caption{Comparison of the performance distribution of Random and KmeansRND regarding different groups of examples on GPT-3.5 and ChatGPT.}
    \label{fig:gpt}
\end{figure}

\subsection{RQ4: The Generalization of Findings}

\subsubsection{Experimental setup}
In this section, we evaluate the generalization of our findings on different LLMs. Apart from Codex, we experiment on two other LLMs including GPT-3.5~\cite{DBLP:conf/nips/BrownMRSKDNSSAA20} and ChatGPT~\cite{ChatGPT}. To validate the finding 1-4,
we experiment with the following combinations of demonstration selection and ordering methods: Random+Random, KmeansRND+Random, UniXcoder+Random, BM-25+Random, and BM-25+Similarity. As for the finding 5
in RQ3, we use BM-25+Similarity as the selection and ordering method and vary the number of demonstration examples from 1 to 128 to validate whether the truncation will lead to performance degradation. Due to the cost limit,
we choose the program synthesis task for evaluation.

We also measure how much improvement could our findings bring by comparing the performance of ICL with a carefully-designed demonstration, ICL with the widely-used demonstration construction method
\cite{DBLP:journals/corr/abs-2210-14179,DBLP:conf/icse-apr/PrennerBR22,DBLP:conf/kbse/Khan022}, and zero-shot ICL. In the carefully-designed demonstration, we use BM-25 and Similarity as demonstration selection and ordering methods and employ four demonstration examples; while for the widely-used baseline demonstration construction method, we use the settings in previous work~\cite{DBLP:journals/corr/abs-2210-14179,DBLP:conf/icse-apr/PrennerBR22,DBLP:conf/kbse/Khan022} and randomly select two demonstration examples from the training set with random order.
As for zero-shot ICL, as illustrated in section~\ref{sec:back_icl}, no demonstration example is used and the model predicts only based on the instruction.

\begin{figure}[t]
    \centering
    \begin{subfigure}[b]{0.25\textwidth}
      \centering
      \includegraphics[width=1\textwidth]{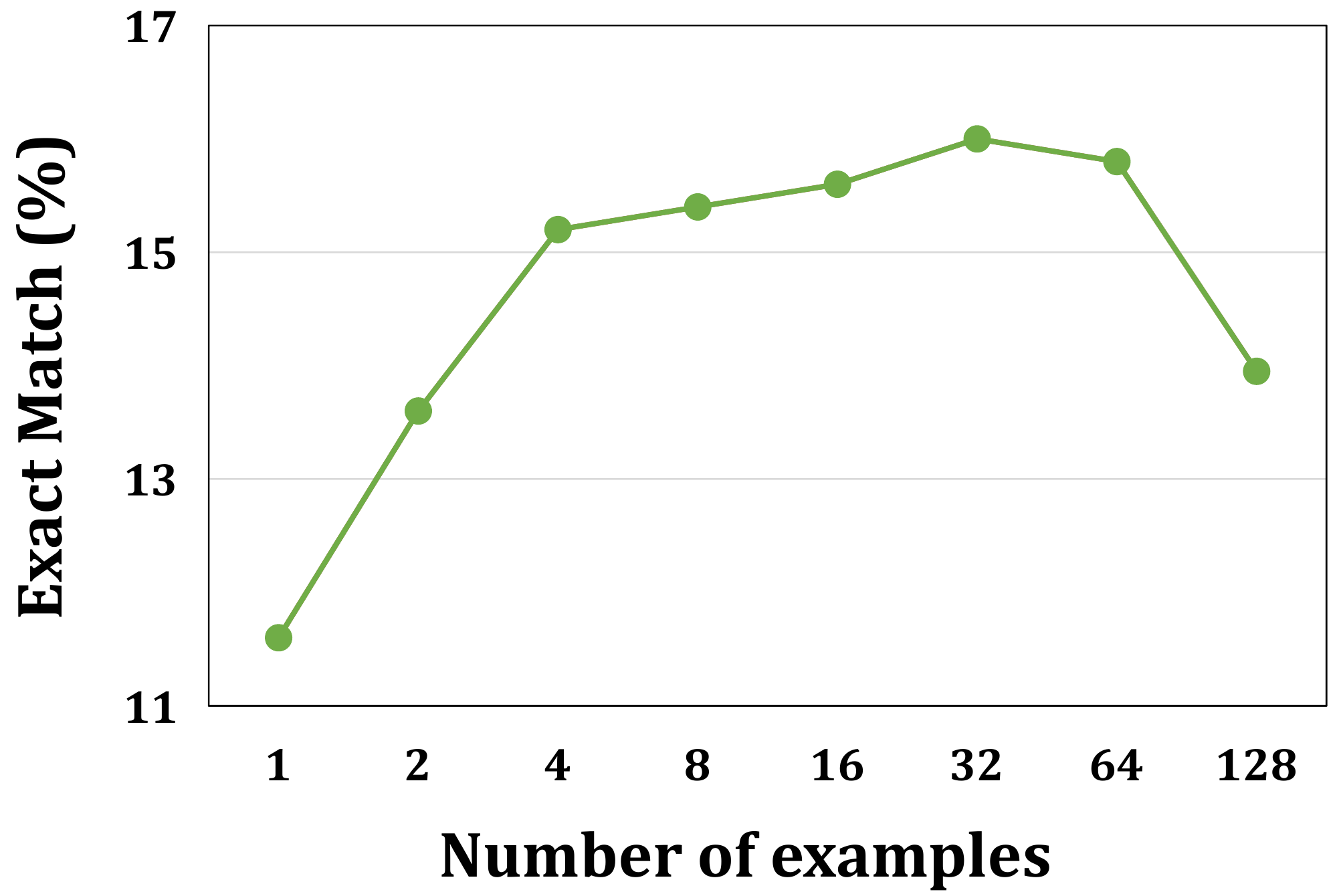}
      \caption{GPT-3.5.}
    \end{subfigure}
    \hfill
    \begin{subfigure}[b]{0.23\textwidth}
      \centering
      \includegraphics[width=1\textwidth]{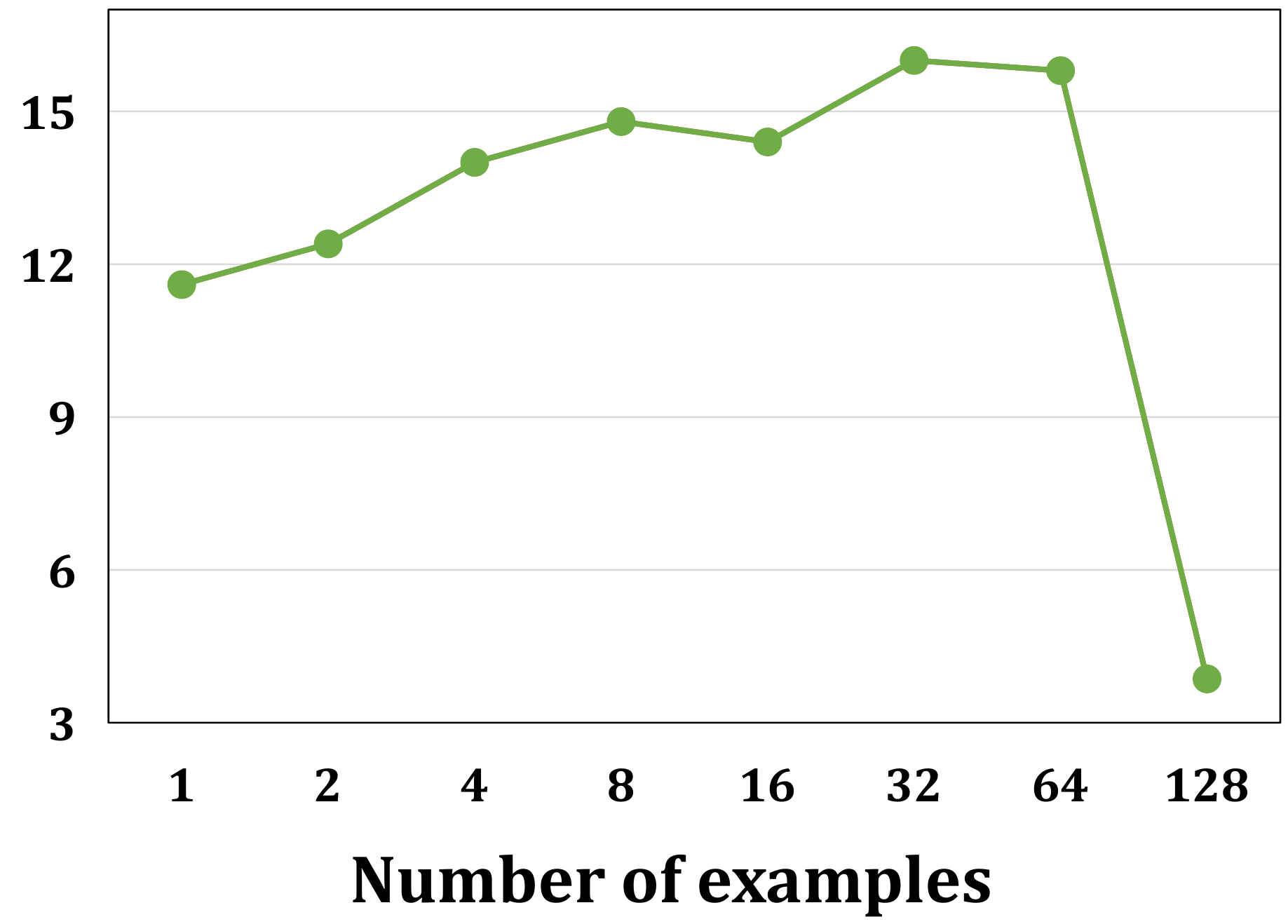}
      \caption{ChatGPT.}
      \end{subfigure}
    \caption{Experimental results of different number of demonstration examples on GPT-3.5 and ChatGPT.}
	\label{fig:generalization}
\end{figure}

\subsubsection{Analysis} 
We present the average results and CV of GPT-3.5 and ChatGPT in Table~\ref{tab:generalization}. In Fig.~\ref{fig:gpt} and Fig.~\ref{fig:generalization}, we present the performance distribution of different groups of examples and the impact of the number of examples on these two LLMs, respectively. The comparison of different demonstrations is shown in Table~\ref{tab:comparison}. Due to the space limitation, we only present the performance on EM and CB and the results on other metrics can be found in our replication package. From these results, we can observe that our findings can also be applied to GPT-3.5 and ChatGPT.

As shown in Table~\ref{tab:generalization} and Fig.~\ref{fig:generalization}, we can observe that KmeansRND+Random not only outperforms Random+Random on the average results, but also has a more stable prediction distribution regarding different groups of examples. Taking GPT-3.5 as an example, 
KmeansRND+Random improves Random+Random by 6.24\% and 10.39\% with respect to CB and EM, respectively. This indicates that diversity is also beneficial for the demonstration construction of these two models \textbf{(finding 1)}. 
Similarly, by comparing BM-25+Random and UniXcoder+Random, we can also find that BM-25 can achieve similar performance and even outperforms UniXcoder on GPT-3.5 by 2.50\% and 4.88\% with respect to CB and EM, respectively. This shows that BM-25 is also a simple and effective demonstration selection method in these two models \textbf{(finding 2)}. 
Besides, on GPT-3.5 and ChatGPT, instance-level demonstrations also consistently outperform task-level demonstrations and achieve lower CV to different orders in general. It indicates that selecting demonstration examples by similarity is also beneficial for these two LLMs \textbf{(finding 3)}. 
As for the impact of example order, we can also find that BM-25+Similarity consistently improves BM-25+Random on all metrics and LLMs, e.g., improving the average EM by 5.56\% and 5.42\% on GPT-3.5 and ChatGPT, respectively \textbf{(finding 4)}. 
As for the impact of numbers, we can observe similar trends on GPT-3.5 and ChatGPT in Fig.~\ref{fig:generalization}, the EM first increases with the number of demonstration examples.
As the number further increases to 128, 25.05\% examples suffer from the truncation problem, resulting in a sudden degradation \textbf{(finding 5)}.

Table~\ref{tab:comparison} shows the comparison of different demonstrations. We can also observe that the performance of zero-shot ICL is very poor on all tasks, which indicates the importance of using demonstration examples to guide the LLM to understand the task. Besides, by comparing the performance of the carefully-designed demonstration with the baseline demonstration, we can find that ChatGPT with a carefully-designed demonstration outperforms the baseline demonstration by at least 10.59\%, 294.57\%, and 51.06\% on code summarization, bug fixing, and program synthesis with respect to BLEU-4, EM, and EM, respectively. The results indicate
the importance of constructing a good demonstration, and the generalizability of the findings.

\section{Discussion}\label{sec:discuss}

\subsection{Implications of Findings}

In this section, we discuss the implications of our work for 
researchers and 
developers.

\textbf{Researchers:}
Our research demonstrates that the performance of few-shot in-context learning is highly dependent on the design of demonstrations. With well-constructed demonstrations, ICL can achieve much better performance. 
Our experimental results also show 
potential research directions 
in the era of LLM and ICL for the code intelligence community. 
Specifically: 

\begin{itemize}
\item 
As shown in the results of RQ1, current state-of-the-art code retrieval models still have a large gap with the Oracle, indicating that these models fail to select examples with the highest semantic similarities. 
Therefore, effective code representation models for zero-shot code-to-code search are worth studying. Besides, designing example selection strategies based on the prior knowledge of each task or the properties of source code 
are also interesting directions that are worth exploring. 

\item Placing similar examples in the back of all examples leads to relatively better performance than random and reverse placings. However, such improvement is not consistent. Therefore, how to automatically design a better ordering method for code intelligence tasks needs to be further investigated.

\item Different from natural language text, the length of a code snippet is 
often much longer. This limits the number of examples in the prompt and could bring large computation and time costs for LLMs. Therefore, incorporating program slicing and reduction techniques into ICL to 
reduce the costs is 
worth investigating.
\end{itemize}

\textbf{Developers:}
In-context learning is a paradigm that allows for learning from a few examples in the prompt without requiring parameter updates. This new approach has also fascinated the language-model-as-a-service community.  Our 
findings indicate that the selection, order, and number of demonstration examples have  significant impacts on 
the performance of ICL for code intelligence tasks. Based on our findings, we conclude the following insights and takeaways for developers to use LLM 
in their work:

\begin{itemize}
\item 
Including demonstration examples in the prompt, which help the model understand the task and guide the output format.

\item 
Using a
retrieval method to select demonstration examples when a labeled training set is available.
For the retrieval methods, consider using BM-25 as it is a simple yet effective method.

\item Improving the diversity of task-level demonstration examples with clustering to obtain more accurate and stable predictions.

\item When arranging the order of demonstration examples, placing similar samples at the end of the list is a good choice in most cases.

\item Using as many demonstration examples as possible, but be mindful of the maximum length limitation to avoid truncation issues. 
To save costs, it is also suggested that four 
examples are used in the demonstration.
\end{itemize}


\subsection{Threats to Validity}

We identify three main threats to validity of our study:

\begin{enumerate}
\item \textbf{Potential data leakage.} In this paper, we conduct experiments by using the API of OpenAI Codex, GPT-3.5, and ChatGPT. However, since they are closed-source models, their parameters and training sets are not publicly available, which raises concerns about potential data leakage. Specifically, there is a possibility that the model has already been trained on the test set 
and merely memorizes the results instead of predicting them. However, we can observe from our experiments that the model's performance in a zero-shot setting is catastrophic, indicating a low probability of direct memorization of the dataset. Moreover, all experiments in our paper were conducted using these models and we use the relative performance improvement to measure the effectiveness of different demonstration construction strategies. Therefore, the findings of our paper remain convincing.

\item \textbf{The selection of tasks.} In this study, we investigate constructions of the demonstration on representative three tasks including code summarization, bug fixing, and program synthesis. 
These tasks cover different types such as Code $\to $ Text, Code+Text $\to $ Code, and Text $\to $ Code. Hence, we believe the finding of our paper can generalize to a wide arrange of code intelligent tasks. In the future, we plan to conduct experiments on other types of tasks such as Code $\to $ Class tasks (e.g., vulnerability detection) and Code $\to $ Code tasks (e.g., code translation).

\item \textbf{The selection of models.} In this paper, we select three LLMs for experiments. Nonetheless, there are other LLMs available, such as CodeGen~\cite{nijkamp2022codegen} and CodeGeeX~\cite{CodeGeeX}. 
In the future, we plan to conduct experiments on a broader range of LLMs to verify the generalizability of our findings.

\item \textbf{The selection of languages.} For each task, we select one popular dataset for evaluation. The datasets of three tasks only contain two programming languages, i.e., Java and Python. In the future, we will validate the effectiveness of demonstration construction methods in other languages.

\end{enumerate}
\section{Related work}\label{sec:related}

\subsection{Pre-trained Models of Code}
Recently, with the development of pre-trained techniques, the pre-trained models of code have been widely used and achieved state-of-the-art performance on various software engineering tasks. One such model is CodeBERT~\cite{DBLP:conf/emnlp/FengGTDFGS0LJZ20}, which is an encoder-only pre-trained model on six programming languages with two self-supervised tasks. 
Another model, CodeT5~\cite{DBLP:conf/emnlp/0034WJH21} is an encoder-decoder pre-trained model following the same architecture as T5. 
CodeGPT~\cite{DBLP:journals/corr/abs-2102-04664} is a decoder-only model that pre-trains on programming languages dataset and has the same architecture as GPT-2. 
PLBART~\cite{DBLP:conf/naacl/AhmadCRC21} uses denoising sequence-to-sequence pretraining for both program understanding and generation purposes. UniXCoder~\cite{DBLP:conf/acl/GuoLDW0022} involves multi-modal contrastive learning and cross-modal generation objective to learn the representation of code fragments. 

Apart from these smaller pre-trained models in academic circles, many pre-trained code models with much larger sizes have been proposed in the industry in recent years. 
Codex~\cite{DBLP:journals/corr/abs-2107-03374} is a large code pre-trained model proposed by OpenAI that supports the service of Copilot. In addition to Codex, the models recently released by OpenAI, such as ChatGPT~\cite{ChatGPT} and GPT-4~\cite{GPT4}, are also pre-trained on source code data and demonstrate impressive programming abilities.  AlphaCode~\cite{DBLP:journals/corr/abs-2203-07814} is trained for generating code for programming competitions like Codeforces, using 715G data and 41B parameters. CodeGen~\cite{nijkamp2022codegen} is a large pre-trained model for multi-turn program synthesis with more than 16B parameters, while CodeGeeX~\cite{CodeGeeX} is a recently proposed open-source multilingual code generation model with 13B parameters.

\subsection{In-context Learning}
Large language models have revolutionized natural language processing (NLP) in recent years. Based on large pre-training data and model sizes, LLMs show impressive emergent abilities that have not been observed in small models~\cite{DBLP:journals/corr/abs-2206-07682}. Brown et al. ~\cite{DBLP:conf/nips/BrownMRSKDNSSAA20} first show that GPT-3 has the ability to learn from a few examples in the context without parameter update. Liu et al.\cite{DBLP:conf/acl-deelio/LiuSZDCC22} first explore selecting the closest neighbors as the in-context examples. Recently, Levy et al.~\cite{DBLP:journals/corr/abs-2212-06800} propose to improve the diversity of in-context examples and achieve better performance on NLP compositional generalization tasks. 
Lu et al.\cite{DBLP:conf/acl/LuBM0S22} find that the order of in-context examples has a large impact on the performance and propose two methods LocalE and GlobalE based on the entropy. 
Recently, a series of work~\cite{DBLP:journals/corr/abs-2201-11903,DBLP:journals/corr/abs-2205-11916} focus on the complex reasoning tasks and propose chain-of-thought prompt by guiding the model to output its reasoning path.  

In addition to NLP, there has been increasing interest in applying in-context learning to code intelligence tasks~\cite{DBLP:journals/corr/abs-2210-14179,DBLP:conf/icse-apr/PrennerBR22,nashidretrieval,DBLP:conf/kbse/Khan022,DBLP:conf/kbse/AhmedD22,DBLP:conf/iclr/PoesiaP00SMG22}. For example, Xia et al.~\cite{DBLP:journals/corr/abs-2210-14179} evaluate the effectiveness of LLMs on program repair. Nashid et al.~\cite{nashidretrieval} propose to use the BM-25 to retrieve similar examples and construct the demonstrations for assert generation and program repair. However, these works mainly focus on the evaluation of LLMs on one or two tasks and do not discuss the construction of in-context demonstrations in-depth. In contrast, our work aims at conducting a systematic study of designing better demonstrations for ICL in code intelligence tasks.
\section{Conclusion and future work}\label{sec:conclusion}
In this paper, we experimentally investigate the impact of different demonstration selection methods, different demonstration ordering methods, and the number of demonstration examples on the performance of in-context learning for code intelligence tasks. Our research demonstrates that a carefully-designed 
demonstration for ICL outperforms simpler demonstrations a lot. We summarize our findings and provide suggestions to help researchers and developers
construct better demonstrations for code intelligence tasks. In the future, we will explore more aspects of source code on the performance of in-context learning such as 
the quality of the code and the naturalness of the code. 
Additionally, we will also further verify our findings on other large language models. Our source code and full experimental results are available at \url{https://github.com/shuzhenggao/ICL4code}.


\bibliographystyle{IEEEtran}
\bibliography{sample-base.bib}

\end{document}